\documentclass[pra,twocolumn,preprintnumbers,amsmath,amssymb,showpacs,
nofootinbib,floatfix]{revtex4}

\usepackage{graphicx,bm,amsmath}

\makeatletter

\def\graphicscale{\twocolumn@sw{0.30}{0.33}}
\def\graphicthreescale{\twocolumn@sw{0.30}{0.33}}

\makeatother

\begin{document}

\title{Interplay between temperature and trap effects 
in one-dimensional lattice systems of bosonic particles}

 \author{Giacomo Ceccarelli, Christian Torrero, Ettore Vicari}
   \affiliation{Dipartimento di Fisica dell'Universit\`a di Pisa and
   I.N.F.N., Sezione di Pisa, Largo Bruno Pontecorvo 2, I-56127 Pisa,
   Italy} \date{August 4, 2011}

\begin{abstract}
We investigate the interplay of temperature and trap effects in cold
particle systems at their quantum critical regime, such as cold
bosonic atoms in optical lattices at the transitions between
Mott-insulator and superfluid phases.  The theoretical framework is
provided by the one-dimensional Bose-Hubbard model in the presence of
an external trapping potential, and the trap-size scaling theory
describing the large trap-size behavior at a quantum critical point.
We present numerical results for the low-temperature behavior of the
particle density and the density-density correlation function at the
Mott transitions, and within the gapless superfluid phase.
\end{abstract}


\maketitle


Many-body phenomena in dilute atomic gases can be investigated in
experiments of cold atoms in optical lattices created by laser-induced
standing waves, such as Mott-Hubbard transitions in bosonic gases,
see, e.g.,
Refs.~\cite{BDZ-08,GBMHS-02,SMSKE-04,PWMMFCSHB-04,KWW-05,HSBBD-06,FWMGB-06,
SPP-07,CFFFI-09}.  An important feature of these experiments is the
presence of a confining potential which traps the particles within a
limited spatial region of the optical lattice.  The theoretical
framework~\cite{JBCGZ-98} is based on the Bose-Hubbard (BH)
model~\cite{FWGF-89}
\begin{eqnarray}
H_{\rm BH} &=& -{J\over 2}
\sum_{\langle ij\rangle}(b_i^\dagger b_j+b_j^\dagger b_i)
+ {U\over2} \sum_i n_i(n_i-1)
\nonumber \\&+&
\mu \sum_i n_i + \sum_i V(r_i) n_i ,
\label{bhm}
\end{eqnarray}
where $\langle ij\rangle$ is the set of nearest-neighbor sites, and
$n_i\equiv b_i^\dagger b_i$ is the particle density operator.  We
assume a power-law space dependence for the external potential which
gives rise to the trap, i.e.
\begin{equation}
V(r)= v^p r^p,
\label{vrp}
\end{equation}
with even $p$.
The trap size is defined as 
\begin{equation}
l\equiv J^{1/p}/v. 
\label{trside}
\end{equation}
Far from the origin the potential $V(r)$ diverges, therefore $\langle
n_i\rangle$ vanishes and the particles are trapped.  Experiments are
usually set up with a harmonic potential, i.e. $p=2$. In the following
we set $J=1$, so that $l=1/v$.

The homogeneous BH model without trap undergoes continuous quantum
transitions between superfluid and Mott-insulator phases, driven by
the chemical potential $\mu$. In one and two dimensions the length
scale of the critical modes diverges with the exponent $\nu=1/2$ and
the low-energy spectrum scales with the dynamic exponent
$z=2$~\cite{FWGF-89,footnote-2d}.  The superfluid phase is gapless; in
one-dimensional systems the low-energy modes are described by a
two-dimensional conformal field theory.

In the presence of a trapping potential, theoretical and experimental
results have shown the coexistence of Mott-insulator and superfluid
regions when varying the total occupancy of the lattice, see, e.g.,
Refs.~\cite{FWMGB-06,JBCGZ-98,BRSRMDT-02,KPS-02,KSDZ-04,PRHD-04,WATB-04,
RM-04,DLVW-05,GKTWB-06,ULR-06,RBRS-09}.  However, at fixed trap size,
the system does not develop a critical behavior with a diverging
length scale~\cite{BRSRMDT-02,WATB-04}.  Criticality can be recovered
in the limit of large trap size, by simultaneously tuning the chemical
potential to the critical values $\mu_c$ of the homogenous system.
This critical regime can be described in the framework of the
trap-size scaling (TSS) theory~\cite{CV-10,CV-09}.  At a quantum
critical point, the large trap-size behavior of the free-energy
density of a $d$-dimensional trapped particle system is expected to
behave as~\cite{CV-10,CV-10-bh}
\begin{equation}
F(\mu,T,l,x) = l^{-\theta(d+z)} 
{\cal F}(\bar{\mu} l^{\theta/\nu},Tl^{\theta z},xl^{-\theta}),
\label{freee}
\end{equation}
where $x$ is the distance from the middle of the trap,
$\bar{\mu}\equiv \mu-\mu_c$ and $\mu_c$ is the critical value of the
chemical potential.
The trap exponent $\theta$ at the Mott transitions is given by
\begin{equation}
\theta= {p\over p+2},
\label{thetaexp}
\end{equation}
which determines how the length scale of the critical modes diverges
with increasing trap size, i.e.  $\xi\sim l^{\theta}$, at the critical
point.  Analogously, the particle density correlator,
\begin{eqnarray}
G(x,y)\equiv \langle n_x n_y \rangle_c \equiv
\langle n_x n_y \rangle - \langle n_x \rangle \langle n_y \rangle,
\label{gndef} 
\end{eqnarray}
is expected to scale as
\begin{eqnarray}
G(x,y)
= l^{-2d\theta} 
{\cal G}(\bar\mu l^{\theta/\nu},Tl^{z\theta},x/l^\theta,y/l^\theta).
\label{gntss} 
\end{eqnarray}
Finite-size effects, when the trap is within a finite box of size $L$,
can be taken into account by adding a further dependence on the
scaling variable $L/l^\theta$~\cite{QSS-10}.

The TSS of the one-dimensional (1D) BH model has been investigated at
zero temperature~\cite{CV-10-bh}.  TSS can be derived analytically in
the low-density regime, at the superfluid-to-vacuum transition, within
the spinless fermion representation of the hard-core limit. The
trap-size dependence turns out to be more subtle in the other critical
regions, when the corresponding homogenous system has a nonzero
filling $f$.  In the presence of the trapping potential, the
expectation value $N$ of the particle-number operator $\hat{N}=\sum_x
n_x$ is finite, and increases as $N\sim l$ with increasing the trap
size $l$ keeping $\mu$ fixed.  Thus, since $[\hat{N},H_{\rm BH}]=0$,
the lowest-energy states at fixed $\mu$, and in particular the ground
state, change when $N$ jumps by 1 with increasing $l$, giving rise to
an infinite number of level crossings~\cite{CV-10-bh}.  At the Mott
transitions with nonzero integer filling, this gives rise to a
modulated TSS: the TSS of the observables is still controlled by the
trap-size exponent $\theta$, but it gets modulated by periodic
functions of the trap size, requiring a minor revision of the TSS
ansatz (\ref{freee}) and (\ref{gntss}).  For example, the
particle-density correlation function at $T=0$ behaves as $G(0,x)=
l^{-2\theta} g(x/l^\theta,\phi)$ where $\phi$ is a phase-like variable
measuring the distance from the closest level crossing.  Note that
this phenomenon persists in the limit $p\to\infty$, which corresponds
to a homogeneous system with open boundary conditions.  Modulations of
the asymptotic power-law behavior is also found in the gapless
superfluid region, with additional multiscaling behaviors.

In this paper we investigate the interplay of temperature and trap
effects in the 1D BH model (\ref{bhm}) at the transitions between
Mott-insulator and superfluid phases. The results are analyzed and
discussed in the theoretical TSS framework, which provides an
effective description of the critical behaviors in the presence of the
trap.  In particular, we address the issue whether the modulation
phenomenon found at $\mu<1$ and $T=0$ persists at finite temperature,
or it is effectively averaged out due to the fact that the level
crossings of the lowest states are expected to give weaker effects at
$T>0$.  This study is of experimental relevance, because quasi 1D
trapped particle systems have been realized in optical lattices, see,
e.g., Refs.~\cite{BDZ-08,PWMMFCSHB-04,KWW-05,CFFFI-09}.

We present numerical results for the 1D BH model, obtained by quantum
Monte Carlo (QMC) simulations at fixed trap size $l$, based on the
stochastic series expansion~\cite{Sandvik-99,SS-02}.  The main
features of the TSS at the Mott transitions are expected to be
universal with respect to the on-site coupling $U$ (provided $U>0$),
including the hard-core (HC) limit $U\to\infty$. Therefore, we
consider the 1D HC BH model~\cite{xxmodel}, which is particularly
convenient because it restricts the values of the site particle number
to $n_x=0,1$, and it is expected to minimize the corrections to the
universal TSS, as verified at $T=0$~\cite{CV-10-bhn}.  In the absence
of the trap, the 1D HC BH model has three phases: the empty state for
$\mu>1$ with $\langle n_i\rangle=0$, which may be seen as a particular
$n=0$ Mott phase, a gapless superfluid phase for $|\mu|< 1$, and a
$n=1$ Mott phase for $\mu<-1$. See, e.g.,
Ref.~\cite{Sachdev-book}. Therefore, there are two quantum transitions
at $\mu=1$ and $\mu=-1$, which share the same critical exponents
$\nu=1/2$ and $z=2$.  We present finite-temperature results for the
particle density and the density-density correlator, at the
transitions between the Mott and superfluid phases, and within the
gapless superfluid phase.  Their temperature and trap-size dependences
appear well described by the simplest TSS ansatz, such as that given
in Eq.~(\ref{gntss}).  The finite-temperature data do not show
evidence of the modulation phenomenon, which appears averaged out,
showing only a plain power-law TSS behavior.

The paper is organized as follows.  In Sec.~\ref{lowd} we study the
temperature and trap-size dependences in the low-density regime, where
we analytically derive the TSS describing temperature and trap
effects, and compare it with QMC simulations at $\mu=1$.
Sec.~\ref{n1mott} presents QMC results at the $n=1$ Mott
transition for $p=2$ and $p\to\infty$.  In Sec.~\ref{superfluidreg} we
investigate the superfluid phase, presenting QMC results at $\mu=0$.
In Sec.~\ref{mueff1} we study the TSS around the spatial point where
the particle density rapidly vanishes when $\mu<1$, which shows a
peculiar scaling behavior effectively controlled by a linear external
potential.  Finally, we draw our conclusions in
Sec.~\ref{conclusions}.  In App.~\ref{qmcsim} we provide some details
on the QMC simulations.

\section{The low-density regime}
\label{lowd}

In order to determine the temperature and trap-size dependence in the
low-density regime of the 1D HC BH model, around $\mu=1$, we consider
the free fermion representation which can be derived by a
Jordan-Wigner transformation, see, e.g., Ref.~\cite{Sachdev-book},
\begin{eqnarray}
&&H_c = \sum_{ij} c_{i}^\dagger h_{ij} c_{j},\label{fermmod}\\
&&h_{ij} = \delta_{ij} - {1\over 2} \delta_{i,j-1} - {1\over 2} \delta_{i,j+1}
+ [\bar{\mu} + V(x_i)] \delta_{ij}, 
\nonumber
\end{eqnarray}
with $\bar{\mu}\equiv\mu-1$.  In the fermion representation the
Hamiltonian can be easily diagonalized by introducing new canonical
fermionic variables $\eta_k=\sum_i \phi_{ki} c_i$, where $\phi$
satisfies the equation $\sum_j h_{ij}\phi_{kj} = \omega_k \phi_{ki}$,
obtaining $H_c = \sum_k \omega_k \eta_k^\dagger \eta_k$.  The ground
state contains all $\eta$-fermions with $\omega_k<0$.

\begin{figure}[tbp]
\includegraphics*[scale=\graphicscale]{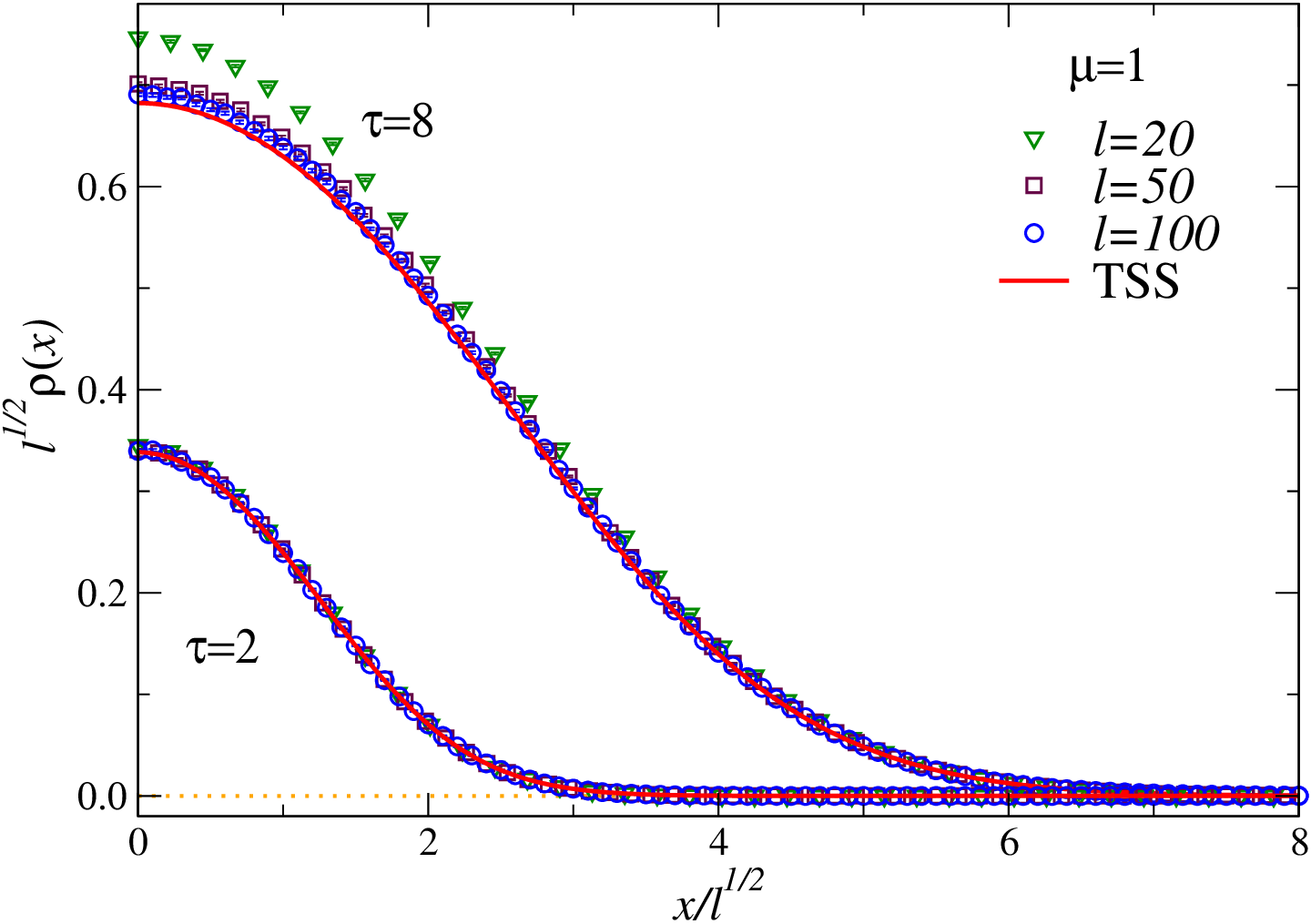}
\includegraphics*[scale=\graphicscale]{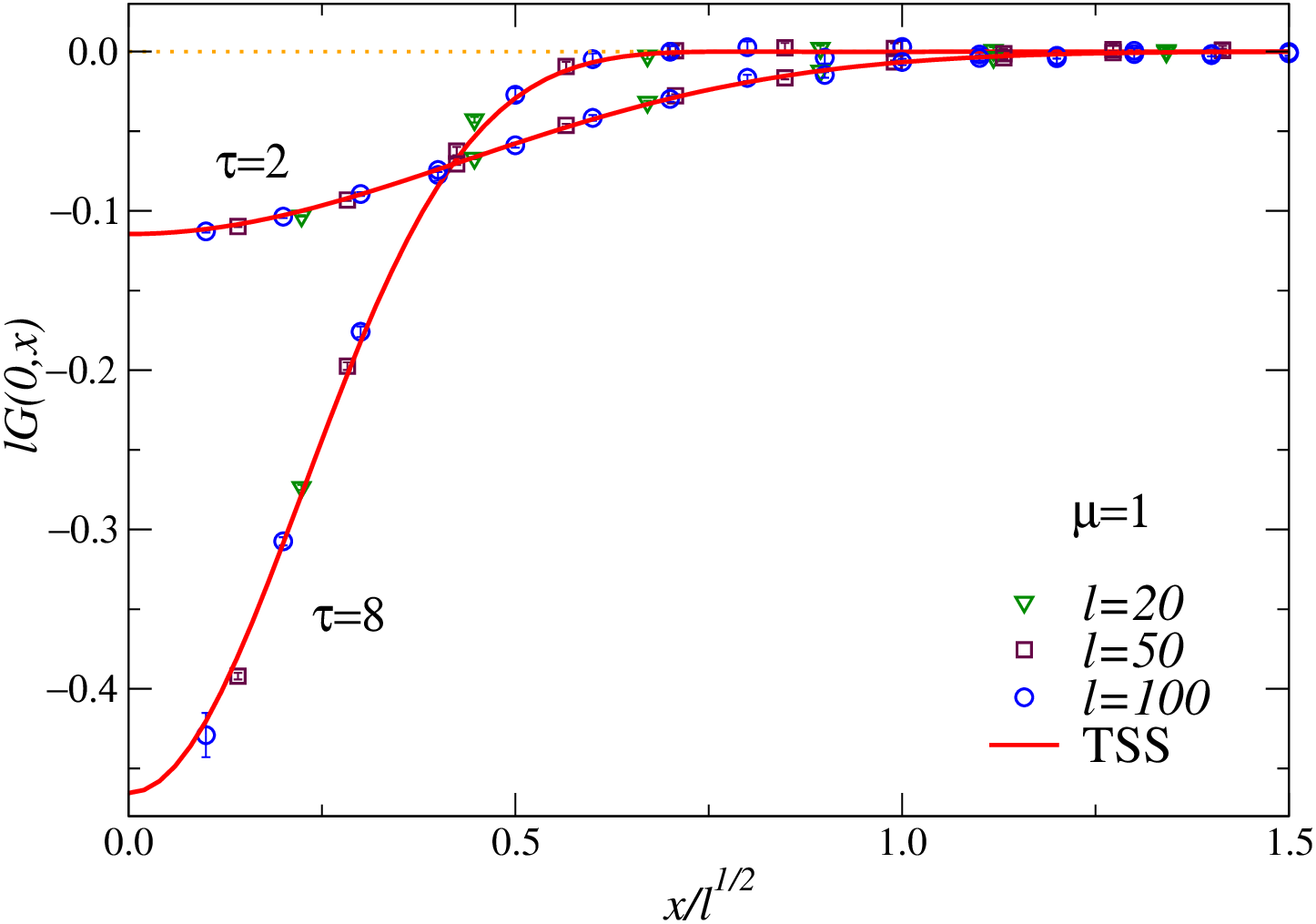}
\caption{(Color online) The particle density (top) and the
density-density correlator (bottom) in the presence of a harmonic
potential, at $\mu=1$ and for some values of the trap size $l$, at
$\tau\equiv Tl=8$ and $\tau=2$.  The full lines show the TSS functions
given by Eqs.~(\ref{dxtl}) and (\ref{gxtl}).  }
\label{demu1}
\end{figure}

A nontrivial TSS limit~\cite{CV-10} around $\mu=1$ is obtained by
introducing the continuum function $\phi_k(x)\equiv \phi_{kx}$, the
rescaled quantities
\begin{eqnarray}
X = x/l^{\theta}, \quad \mu_r = \bar{\mu} l^{2\theta},
\quad \Omega_k = \omega_k l^{2\theta},
\label{resca}
\end{eqnarray}
where $\theta$ is the trap exponent, cf. Eq.~(\ref{thetaexp}), and
neglecting terms which are suppressed in the large-$l$ limit.  This
leads to the Schr\"odinger-like equation
\begin{equation}
\left( - {1\over 2} {d^2 \over dX^2} + X^p \right)\varphi_k(X) =
e_k \varphi_k(X),
\label{trapscaleqxx}
\end{equation}
where $e_k \equiv \Omega_k - \mu_r$ and $\varphi_k(X)\equiv
\phi_k(l^{\theta} X)$.

\begin{figure}[tbp]
\includegraphics*[scale=\graphicscale]{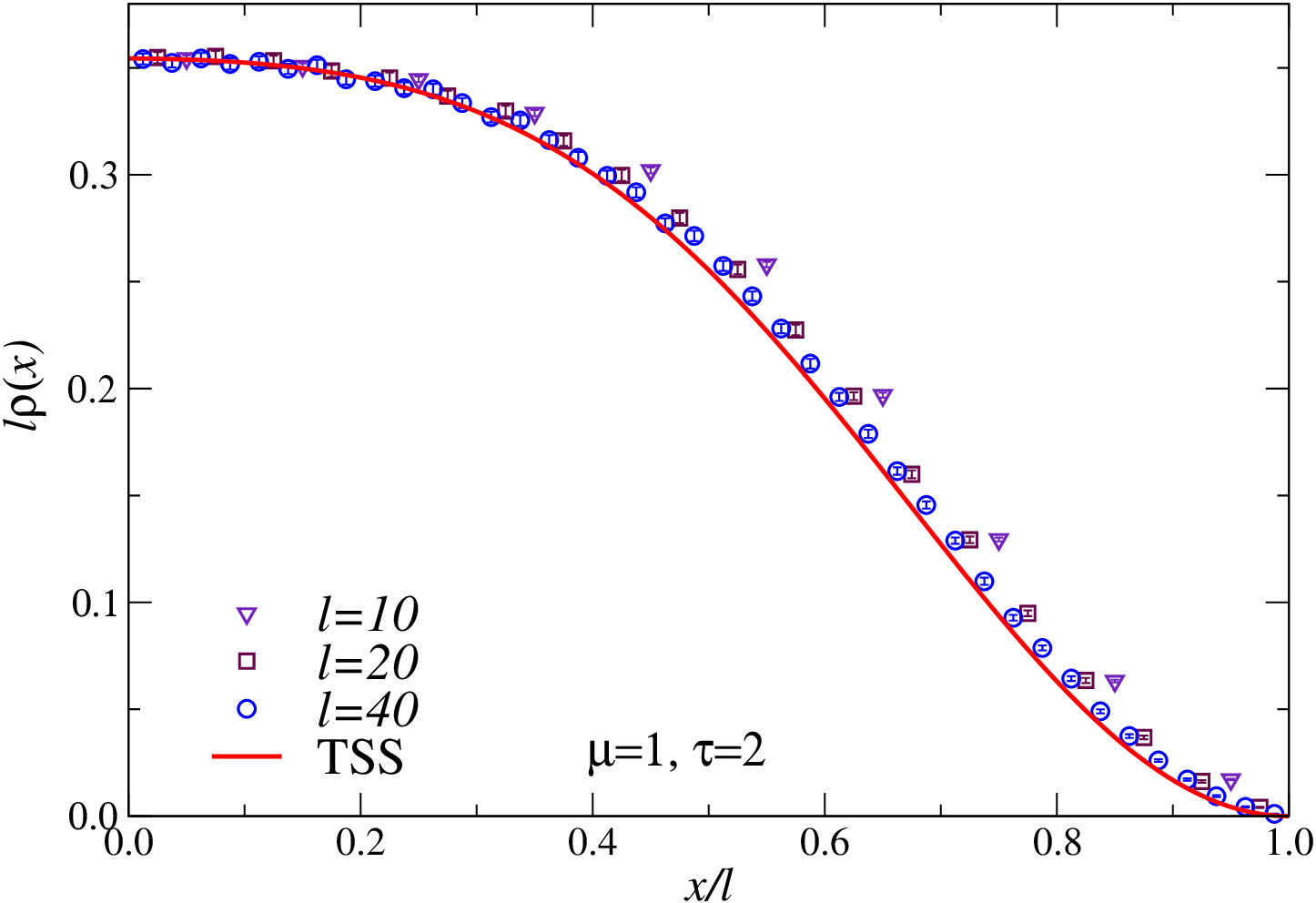}
\includegraphics*[scale=\graphicscale]{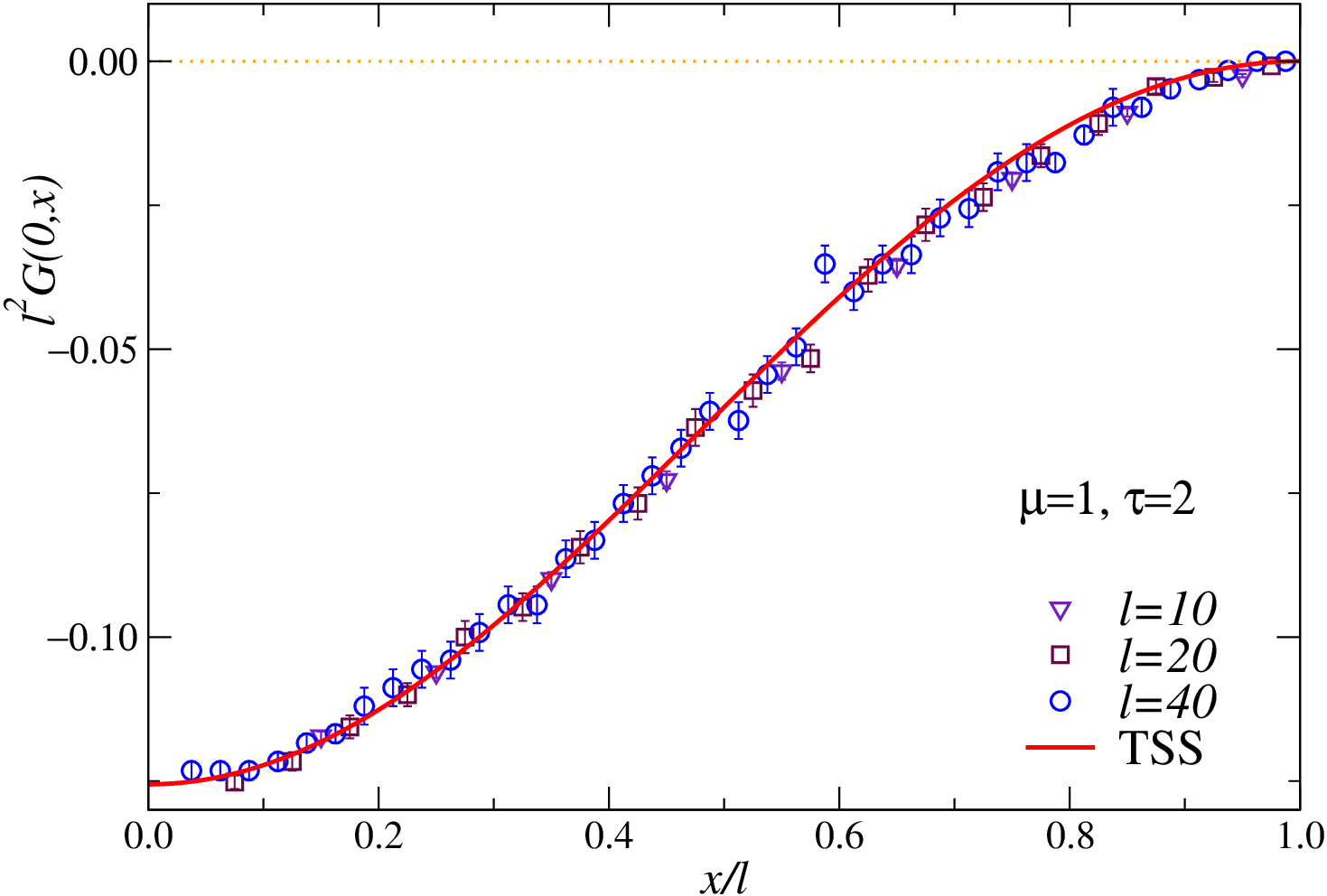}
\caption{(Color online) Results for the particle density (top) and the
density-density correlator (bottom) at $\mu=1$ and $p\to\infty$, for some
values of $l$, keeping $\tau\equiv Tl^2=2$ fixed.  The full
line shows the TSS functions given by Eqs.~(\ref{dxtl}) and
(\ref{gxtl}).  }
\label{mu1tl2pi}
\end{figure}

In the case of a harmonic potential, i.e. $p=2$, we have $\theta=1/2$,
and
\begin{eqnarray}
&&e_k \equiv \Omega_k - \mu_r
= 2^{1/2}(k + 1/2), \quad k\ge 0, \label{eq:p2eig}\\
&&\varphi_k(X) = {2^{1/8} H_k(2^{1/4}X)\over
 \pi^{1/4} 2^{k/2} (k!)^{1/2}} \, \exp(-X^2/\sqrt{2}), 
 \nonumber 
\end{eqnarray}
where $H_k(x)$ are Hermite's polynomials.  For $p\to\infty$, Eq.\
(\ref{trapscaleqxx}) becomes equivalent to the Schr\"odinger equation
of a free particle in a box of size $L=2l$ with boundary conditions
$\varphi(-1)=\varphi(1)=0$, thus $\theta=1$ and
\begin{eqnarray}
&&e_k = {\pi^2\over 8} (k+1)^2, \qquad k\ge 0,
\label{eq:pinfeig}\\
&&\varphi_k(X) = \sin\left[{\pi\over 2} (k+1) (X+1)\right].
\nonumber
\end{eqnarray}

Since
\begin{equation}
n_x\equiv b_x^\dagger b_x=c_x^\dagger c_x,
\label{density}
\end{equation}
the particle density of bosons and its correlation function
(\ref{gndef}) are equal to those of the fermions in the quadratic
Hamiltonian (\ref{fermmod}).  The fermion two-point function is given
by
\begin{eqnarray}
\langle c_x^\dagger c_y \rangle = \sum_{ab}\phi_{ax}\phi_{by} 
\langle \eta_a^\dagger \eta_b \rangle 
= \sum_{k=0}^\infty {\phi_{kx} \phi_{ky}\over 1 + {\rm exp}(\omega_k/T)}.
\label{extwop}
\end{eqnarray}
Its TSS limit can be written in terms of the eigensolutions
of the Schr\"odinger-like equation (\ref{trapscaleqxx}), as
\begin{equation}
\langle c_x^\dagger c_y\rangle =
l^{-\theta} \sum_{k=0}^\infty  {\varphi_k(X) \varphi_k(Y)\over   
1 + {\rm exp}[(e_k + \mu_r)/\tau)} ,
\label{twopfe}
\end{equation}
where we have introduced the scaling variable 
\begin{equation}
\tau\equiv  Tl^{z\theta}=Tl^{2\theta}.
\label{taudef}
\end{equation}
Then, straightforward calculations allow us to extend the $T=0$
results of Ref.~\cite{CV-10-bh}, to allow for the temperature. 
We obtain
\begin{eqnarray}
&&\rho(x)\equiv \langle n_x \rangle =
l^{-\theta} {\cal D}(\bar\mu l^{2\theta},Tl^{2\theta},x/l^\theta),
\label{rhoxmu1} \\
&&{\cal D}(\mu_r,\tau,X) = \sum_{k=0}^\infty {\varphi_k(X)^2\over 
1 + {\rm exp}[(e_k + \mu_r)/\tau]}, \label{dxtl} 
\end{eqnarray}
and
\begin{eqnarray}
&& G(x,y)=
l^{-2\theta} {\cal G}(\bar\mu l^{2\theta},Tl^{2\theta},
x/l^\theta,y/l^\theta), \qquad
\label{gxmu1} \\
&&{\cal G}(\mu_r,\tau,X,Y) = -
\left[ \sum_{k=0}^\infty {\varphi_k(X)\varphi_k(Y)\over 
1 + {\rm exp}[(e_k + \mu_r)/\tau]} \right]^2,\qquad
 \label{gxtl} 
\end{eqnarray}
which holds for $x\neq y$.  Note that $G(x,y)<0$ for $x\neq y$, but
$G(x,x)\equiv \langle n_x^2 \rangle - \langle n_x\rangle^2>0$, indeed
$\sum_{xy} G(x,y)>0$.  The above scaling functions describe the
asymptotic large trap-size behavior; corrections are suppressed by a
further $l^{2\theta}$ power.  Practically exact results for the TSS
functions of the harmonic potential and the hard-wall limit, for any
$\mu_r$ and $\tau$, can be easily obtained using Eqs.~(\ref{eq:p2eig})
and (\ref{eq:pinfeig}), because the series are rapidly converging.

In Fig.~\ref{demu1} we show QMC results for the harmonic potential at
$\mu=1$, for some values of the trap size, keeping $\tau\equiv Tl$
fixed at $\tau=2$ and $\tau=8$.  They clearly appear to approach the
TSS given by Eqs.~(\ref{dxtl}) and (\ref{gxtl}).  Analogous results
are obtained for the hard-wall limit, which is equivalent to a
homogeneous system of size $L=2l$ with open boundary conditions.
Since $\theta=1$ in this case, TSS requires keeping $\tau\equiv Tl^2$
fixed.  Results for $\tau=2$ are shown in Fig.~\ref{mu1tl2pi}.

The TSS functions in the low-density regime are expected to be
universal with respect to the on-site coupling $U$, provided that
$U>0$. As shown by the calculations at $T=0$ reported in
Ref.~\cite{CV-10-bhn}, finite values of $U$ give rise to
$O(l^{-\theta})$ corrections, which decay more slowly than the
$O(l^{-2\theta})$ leading corrections in the HC limit.

\section{TSS at the $n=1$ Mott transition}
\label{n1mott}

We recall that the quantum critical behavior around $\mu=-1$ of the
homogeneous HC BH model without trap is essentially analogous to that
at $\mu=1$, because of the invariance under the particle-hole
exchange.  At the $n=1$ Mott-insulator to superfluid quantum
transition, the critical exponents $\nu$ and $z$ and the trap-size
exponent $\theta$ are the same as those at $\mu=1$.  However, the
particle-hole symmetry does not hold in the presence of the trapping
potential, and the asymptotic trap-size dependence at $T=0$ appears
more complicated at the $n=1$ Mott transition~\cite{CV-10-bh}.

In the 1D HC BH model with a trapping potential and $\mu<1$, the
ground-state particle density turns out to approach its local density
approximation (LDA) in the large-$l$ limit, with corrections that are
suppressed by powers of the trap size and present a nontrivial TSS
behaviour~\cite{CV-10-bh}, see also below.  Within the LDA, the
particle density at the spatial coordinate $x$ equals the particle
density of the homogeneous system at the effective chemical potential
\begin{equation}
\mu_{\rm eff}(x) \equiv \mu + \left({x/l}\right)^p.
\label{mueff}
\end{equation}
The LDA of the particle density reads
$\langle n_x \rangle_{\rm lda} \equiv \rho_{\rm lda}(x/l)$, where
\begin{equation}
\rho_{\rm lda}(x/l) = 
\kern-10pt \quad\left\{
\begin{array}{l@{\ \ }l@{\ \ }l}
0 & {\rm for} & \mu_{\rm eff}(x) > 1, \\
(1/\pi)\arccos\mu_{\rm eff}(x) &
    {\rm for} & -1 \le \mu_{\rm eff}(x) \le 1, \\
1 & {\rm for} & \mu_{\rm eff}(x) < -1. \\
\end{array} \right.
\label{nxlda}
\end{equation}

The {\em thermodynamic} limit at fixed $\mu$ corresponds to the limit
$N,l\to\infty$ keeping the ratio $N/l$ fixed.  At $T=0$ we have
\begin{equation}
N \equiv \langle \sum_i b_i^\dagger b_i  
\rangle = \tilde{\rho}(\mu) l + O(1),\label{cmudef}
\end{equation}
where $\tilde{\rho}(\mu) = 2 \int_0^\infty \rho_{\rm lda}(y) \,{\rm d}
y$.  Eq.~(\ref{cmudef}) provides the correspondence between the ratio
$N/l$ and $\mu$ at $T=0$.  In particular, when
$0<N/l<\tilde{\rho}(-1)$ the system is essentially in the superfluid
phase, while for $N/l>\tilde{\rho}(-1)$ the $n=1$ Mott phase appears
around the center of the trap.

\begin{figure}[tbp]
\includegraphics*[scale=\graphicscale]{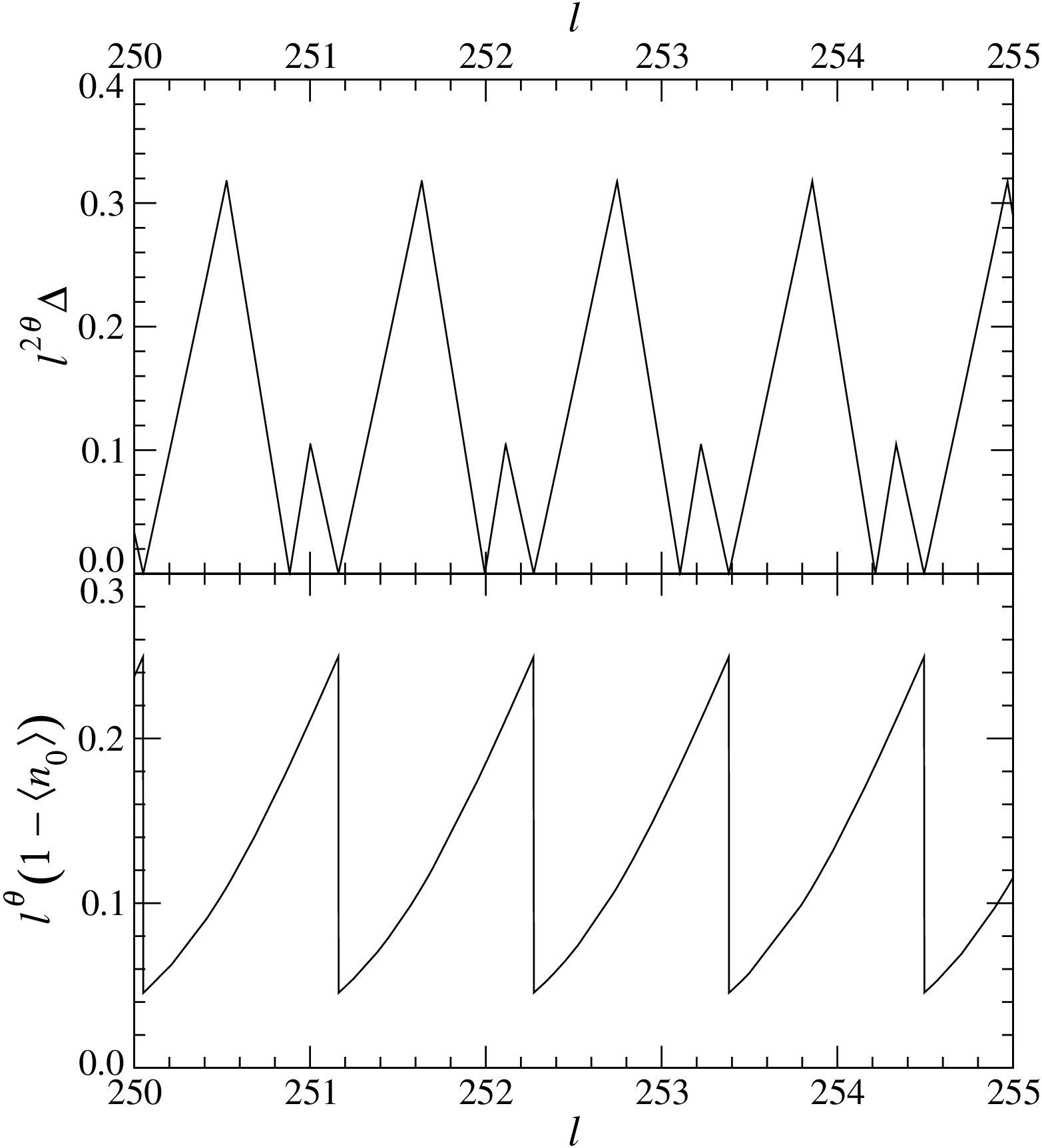} 
\caption{(Color online) The rescaled difference of the lowest energy
levels $l^{2\theta}\Delta$ (above) and the rescaled particle density
$l^{\theta}(1-\langle n_0\rangle)$ (below) at the center of the trap
vs.\ $l$ in the middle of the trap for $p=2$ and at $\mu=-1$.  }
\label{gapanddensityp2}
\end{figure}

The $T=0$ trap-size dependence around $\mu=-1$ is characterized by an
infinite number of level crossings of the two lowest energy states, at
$l_0^{(k)}$ with $k=1,2,3,...$ and $l_0^{(k+1)}>l_0^{(k)}$.  As
suggested by Fig.~\ref{gapanddensityp2}, the rescaled energy
difference $l^{2\theta}\Delta$ of the two lowest states shows a
periodic structure asymptotically for large $l$, as a function of
$l$~\cite{CV-10-bh}.  In particular, in the case of a trap centered at
the middle site of the lattice and in the large-$l$ limit, the
interval
\begin{equation}
P_l \equiv  l_0^{(2k+2)} - l_0^{(2k)},
\label{pldef}
\end{equation}
between two even zeroes of such difference approaches a constant
value, $P_l^* \cong1.11072073$ for $p=2$ and $P_l^*=1$ for
$p\to\infty$.  This gives rise to a peculiar modulation of the
amplitudes of the power-law behaviors of the observables. For example,
the density at the center of the trap behaves as
\begin{equation}
\langle n_0\rangle - 1 \approx  l^{-\theta}  f_{0}(\phi), 
\label{n0scalm-1}
\end{equation}
where $\phi$ is the phase-like variable
\begin{equation}
\phi = {l-l_0^{(2k)}\over  l_0^{(2k+2)}-l_0^{(2k)}}, 
\qquad l_0^{(2k)}\le l < l_0^{(2k+2)},
\label{phildef}
\end{equation}
which measures the distance from the closest even level crossing.
Thus, the amplitudes of the power laws predicted by the TSS show a
periodic modulation, see Ref.~\cite{CV-10-bh} for details.  Moreover,
the spatial dependence of the particle density and its correlator
turns out to be described by the following scaling behavior at large
trap size
\begin{equation}
\rho(x) \equiv \langle n_x\rangle \approx  \rho_{\rm lda}(x/l)
+ l^{-\theta}  f(X,\phi), 
\label{nxscalm-1}
\end{equation}
where $X=x/l^\theta$, and
\begin{equation}
G(0,x) \equiv
\langle n_0 n_x\rangle_c =  l^{-2\theta}  g(X,\phi).
\label{gnxm1}
\end{equation}
An interesting question is whether this scenario persists at finite
temperature, where the effects of the level crossings of the lowest
states are expected to be weaker.

\begin{figure}[tbp]
\includegraphics*[scale=\graphicscale]{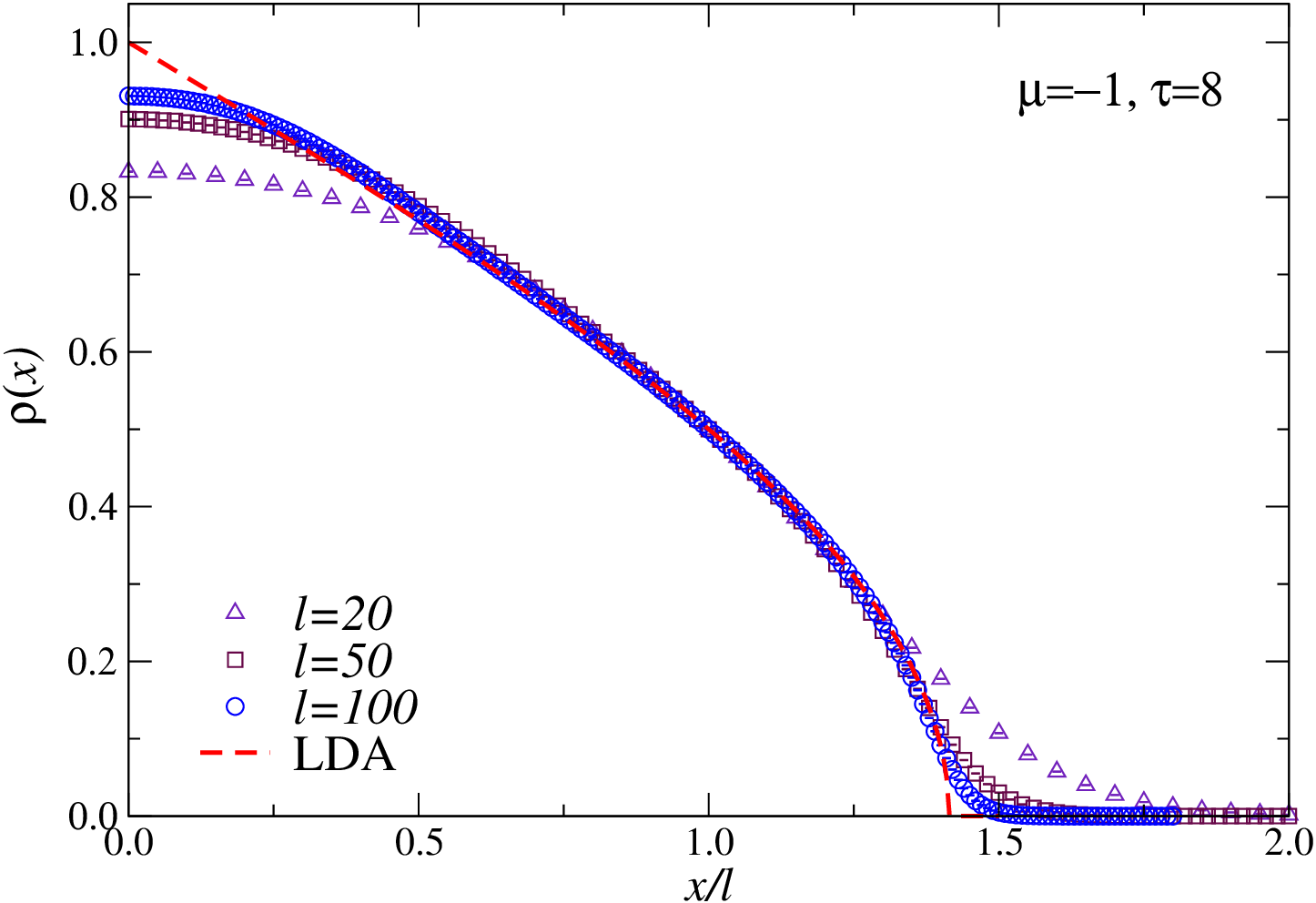}
\includegraphics*[scale=\graphicscale]{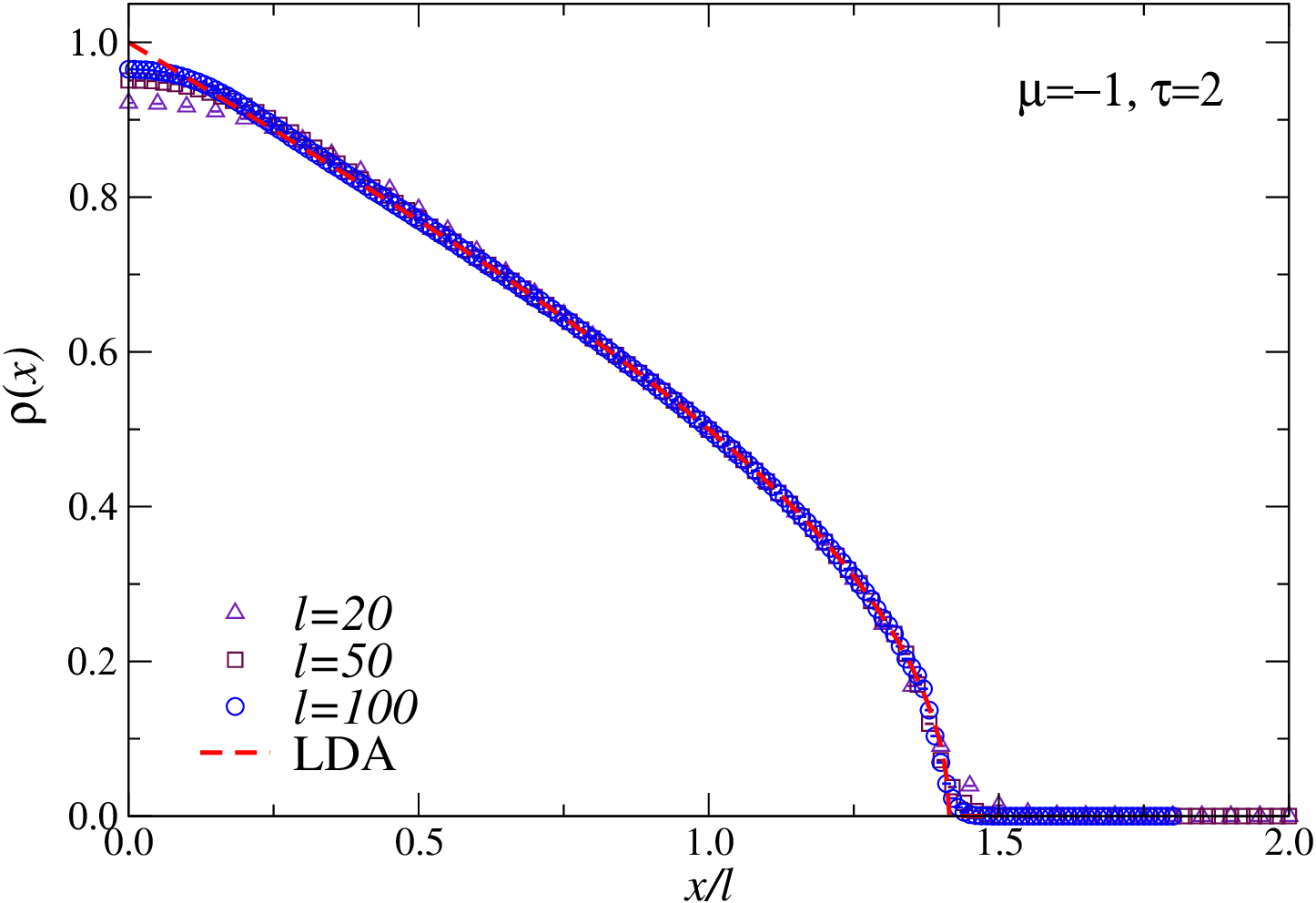}
\caption{(Color online) The particle density at $\mu=-1$ for
$\tau\equiv Tl=8$ (top) and $\tau=2$ (bottom).  The full lines
represent the LDA approximation (\ref{nxlda}).  }
\label{rmum1}
\end{figure}

\begin{figure}[tbp]
\includegraphics*[scale=\graphicscale]{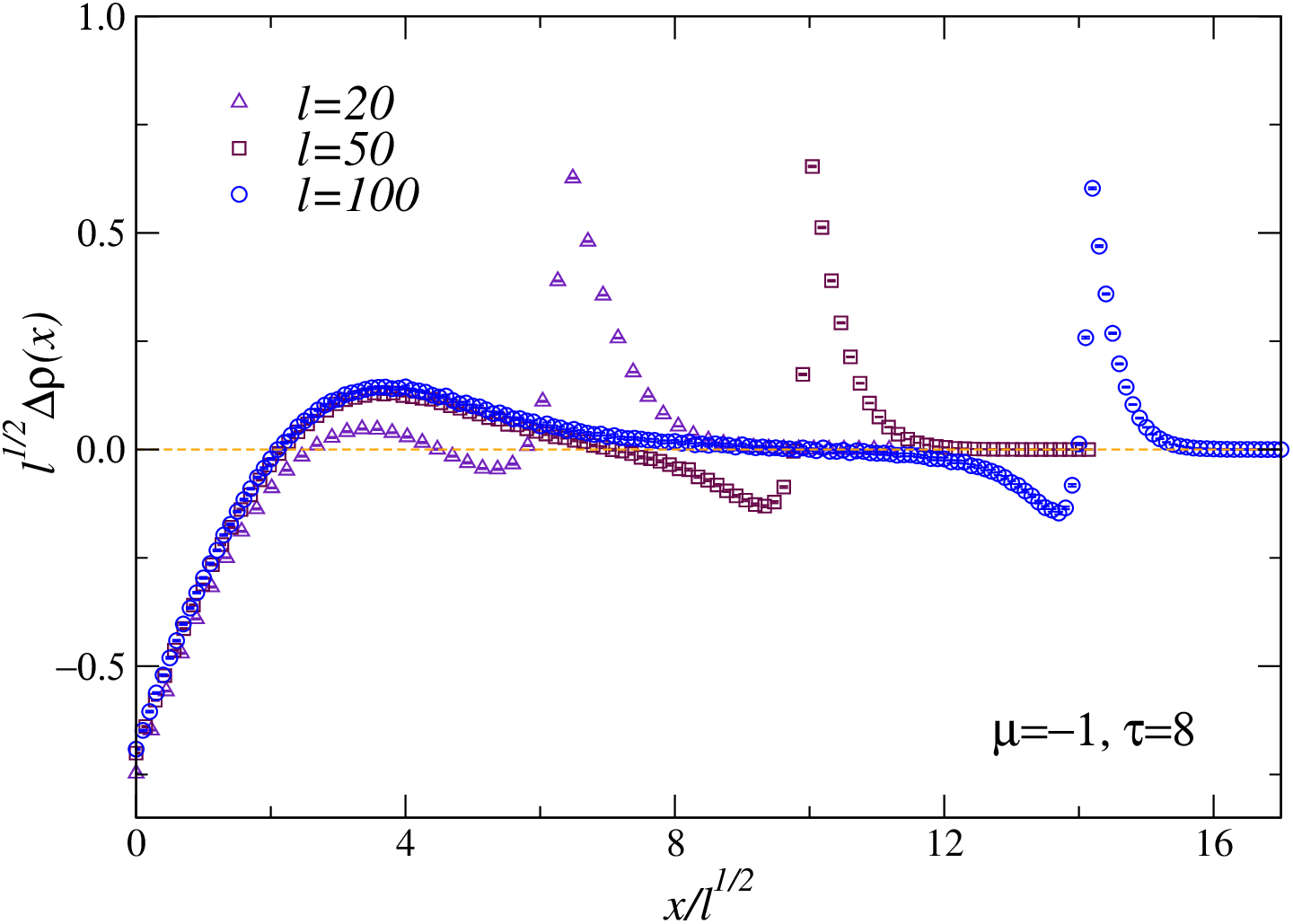}
\includegraphics*[scale=\graphicscale]{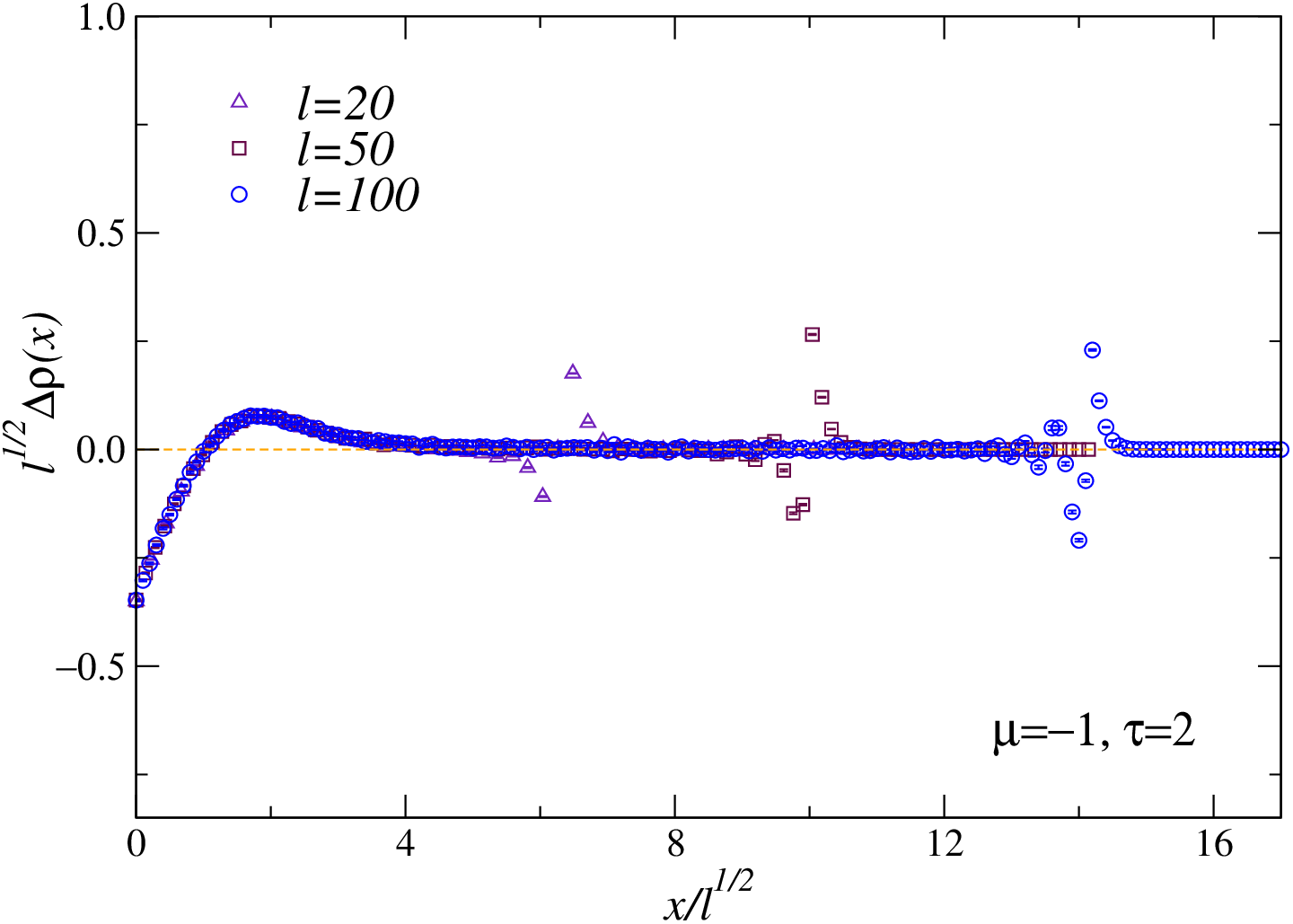}
\caption{(Color online) Scaling of the subtracted particle density at
$\mu=-1$ for $\tau=8$ (top) and $\tau=2$ (bottom).  }
\label{rmum1sub}
\end{figure}

\begin{figure}[tbp]
\includegraphics*[scale=\graphicscale]{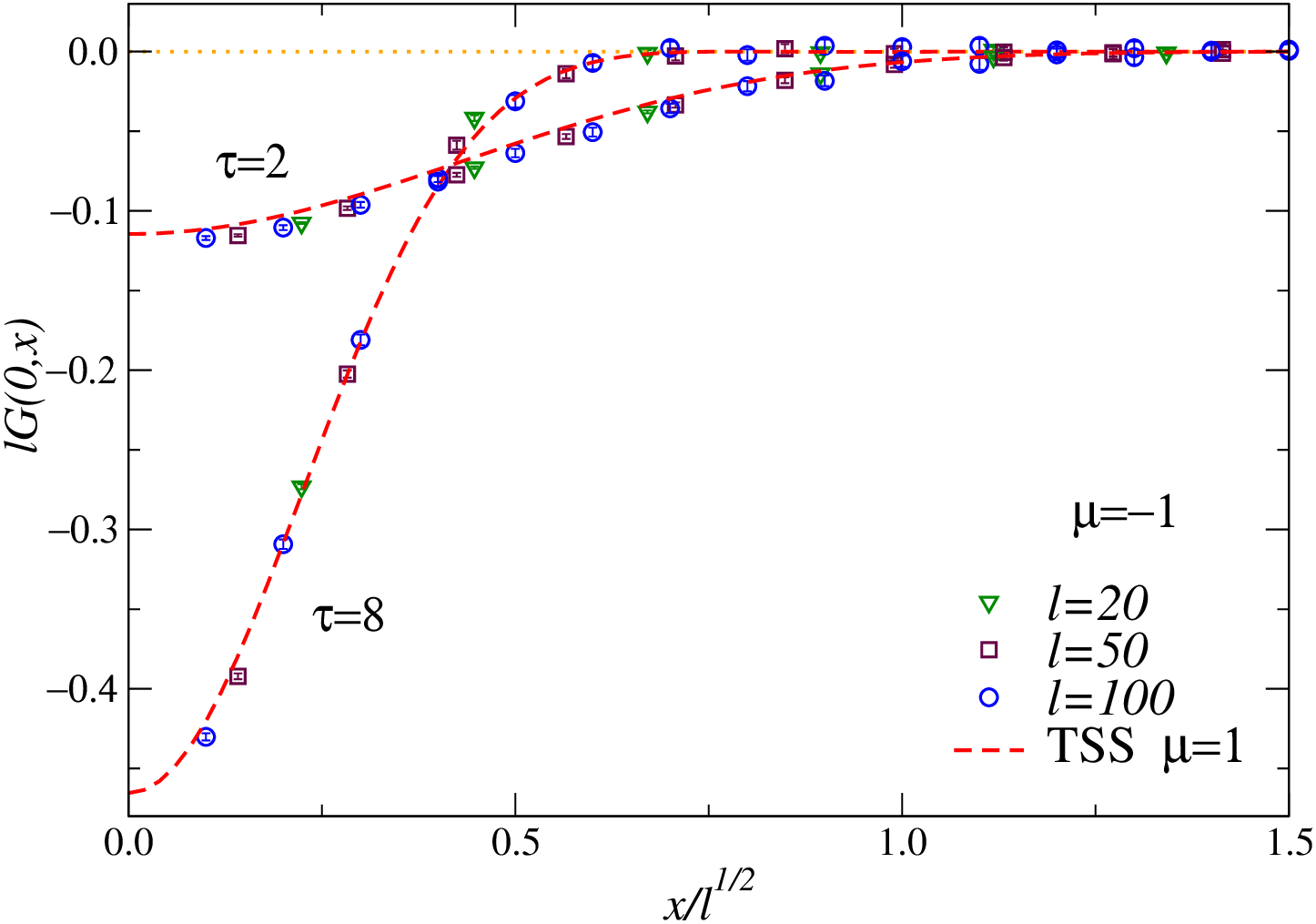}
\caption{(Color online) The density correlation at $\mu=-1$ for values
of the trap size $l$, at $\tau\equiv Tl=8$ and $\tau=2$.  The dashed
lines show the TSS at $\mu=1$.  }
\label{gmum1}
\end{figure}

We investigate this issue by comparing QMC data at various increasing
values of the trap size keeping fixed the scaling quantity $\tau=Tl$.
Fig.~\ref{rmum1} shows results for the particle density, which appears
to approach the LDA approximation (\ref{nxlda}) with increasing $l$,
for $\tau=2$ and $\tau=8$.  This should not be surprising, given that
the LDA approximation is asymptotically approached at $T=0$, and the
TSS is performed by rescaling $T$ as $T=\tau/l$.  Fig.~\ref{rmum1sub}
shows the scaling of the subtracted particle density
\begin{equation}
\Delta\rho(x) \equiv \rho(x) - \rho_{\rm lda}(x/l). 
\label{delrhodef}
\end{equation}
Its behavior around the trap appears to
behave as
\begin{equation}
\Delta\rho(x)  \approx  l^{-\theta}  {\cal D}(X,\tau), 
\label{scadata}
\end{equation}
without evidences of modulations.  Strong deviations from scaling are
observed at large distance from the center of the trap, in a region
corresponding to $x\approx \sqrt{2} l$, which disappears in the TSS
limit, because it moves away with increasing $l$ when the data are
plotted versus the rescaled distance $X\equiv x/l^\theta$ with
$\theta<1$.  As we shall see in Sec.~\ref{mueff1}, the behavior around
$x\approx\sqrt{2}$, where $\mu_{\rm eff}(x)\approx 1$, shows another
scaling behavior, effectively controlled by an external linear
potential.

\begin{figure}[tbp]
\includegraphics*[scale=\graphicscale]{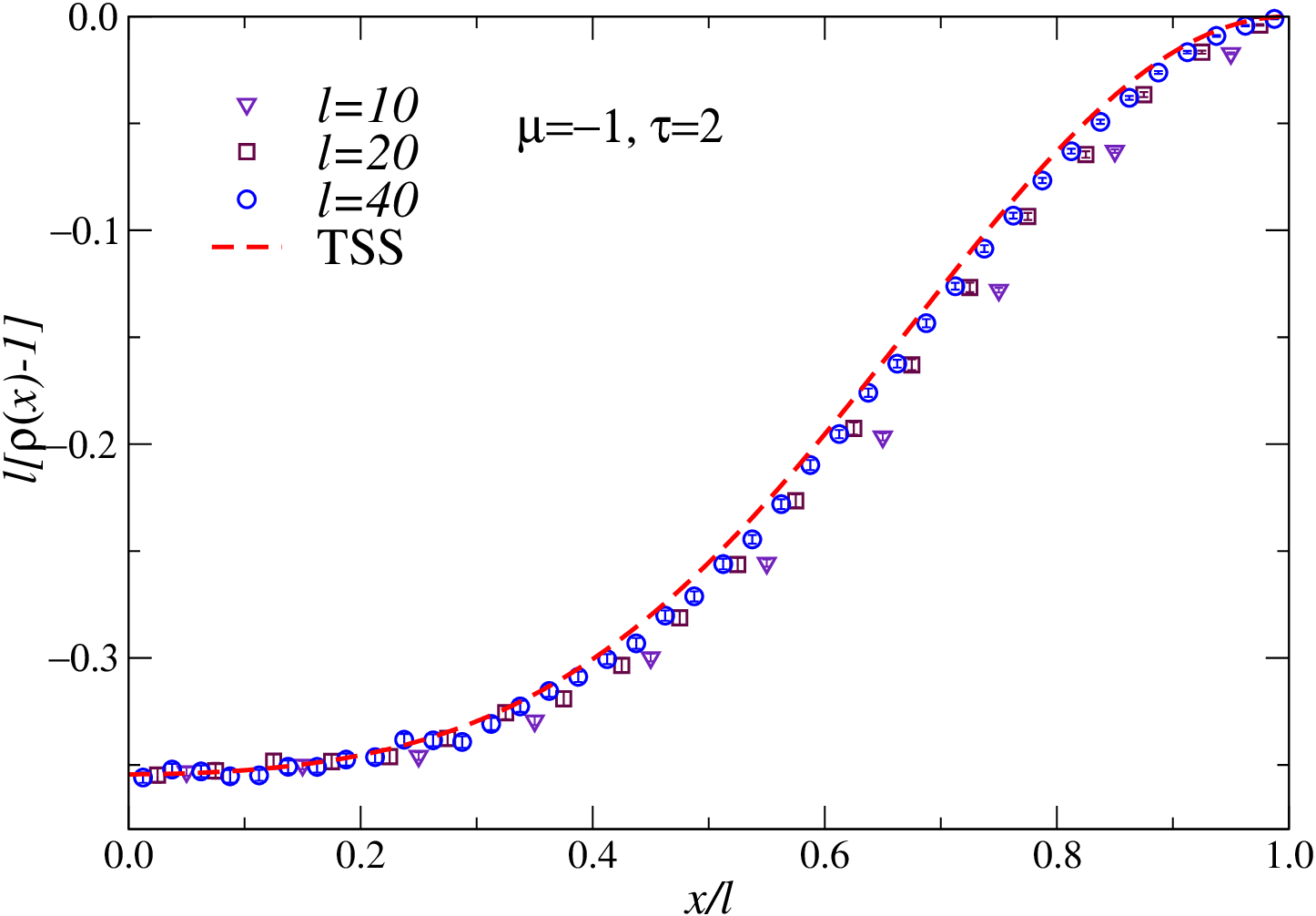}
\includegraphics*[scale=\graphicscale]{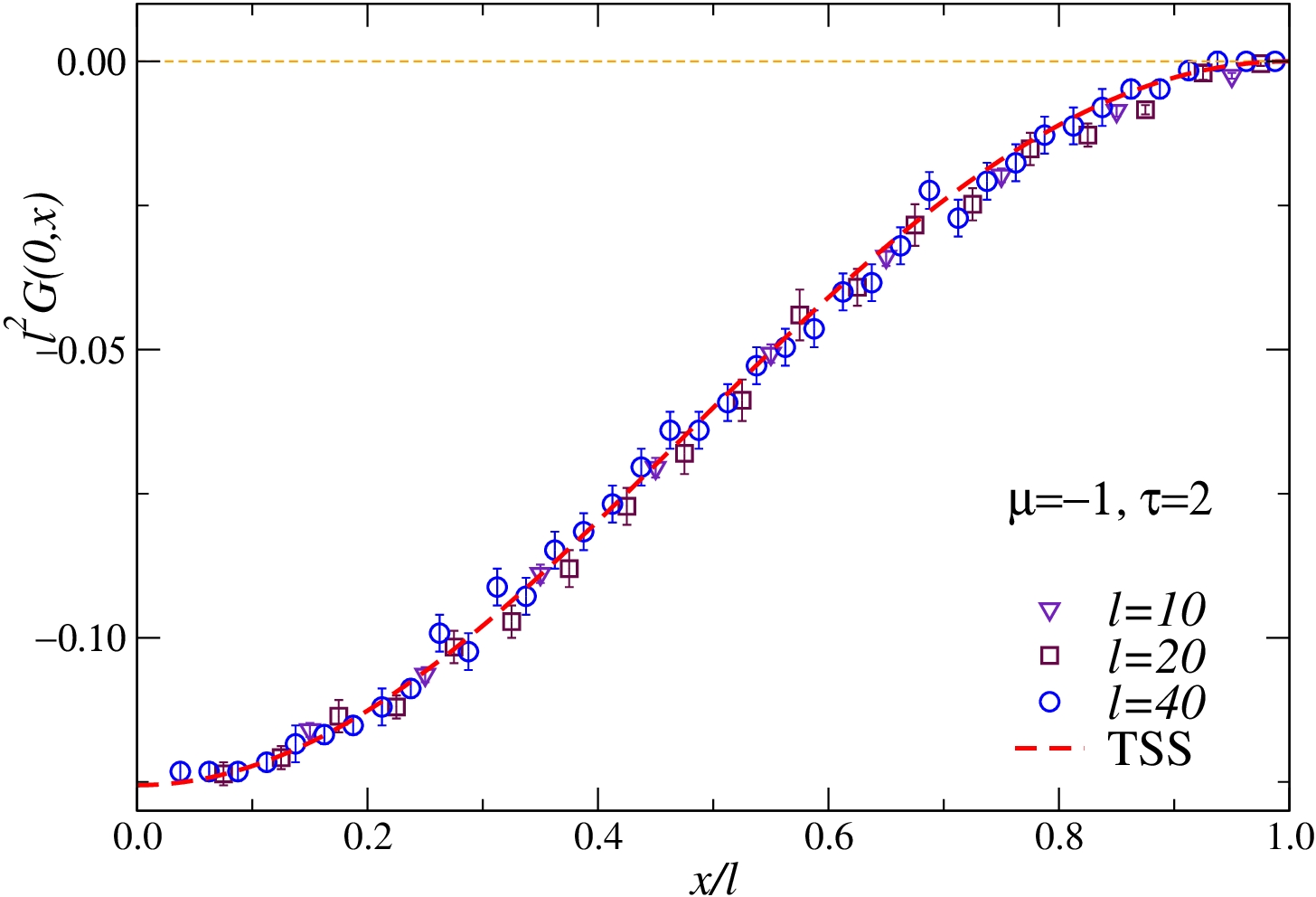}
\caption{(Color online) Plots of $l[\rho(x)-1]$ (top) and $l^2G(0,x)$
(bottom) for $p\to\infty$, at $\mu=-1$ and $\tau\equiv Tl^2=2$.  The
dashed lines show the TSS at $\mu=1$.  }
\label{mum1tl2pi}
\end{figure}

Some results for the particle-density correlation $G(0,x)$, between
the center of the trap and the point $x$, are reported in
Fig.~\ref{gmum1}.  They show a nice scaling compatible with
\begin{equation}
G(0,x) = l^{-1} {\cal G}(X,\tau),
\label{goxsc}
\end{equation}
again without evidence of modulations. Interestingly, the data appear
to approach the same TSS curves found at $\mu=1$ for the same values
of $\tau=Tl$, providing an evidence of universality between the TSS at
the $n=0$ and $n=1$ Mott transitions, at least for the connected
correlation functions.  Analogous results are found in the limit
$p\to\infty$, as shown by the data reported in Fig.~\ref{mum1tl2pi}.

We conclude noting that the apparent absence of modulations indicate
that the temperature averages out these effects, so that the
temperature and trap-size dependences are well described by the TSS
ansatz (\ref{freee}) and (\ref{gntss}), in analogy with the low-density
regime. Of course, the modulation phenomenon must be somehow recovered
in the limit $\tau\to 0$.

\section{TSS within the superfluid phase}
\label{superfluidreg}

\begin{figure}[tbp]
\begin{center}
\includegraphics*[scale=\graphicscale]{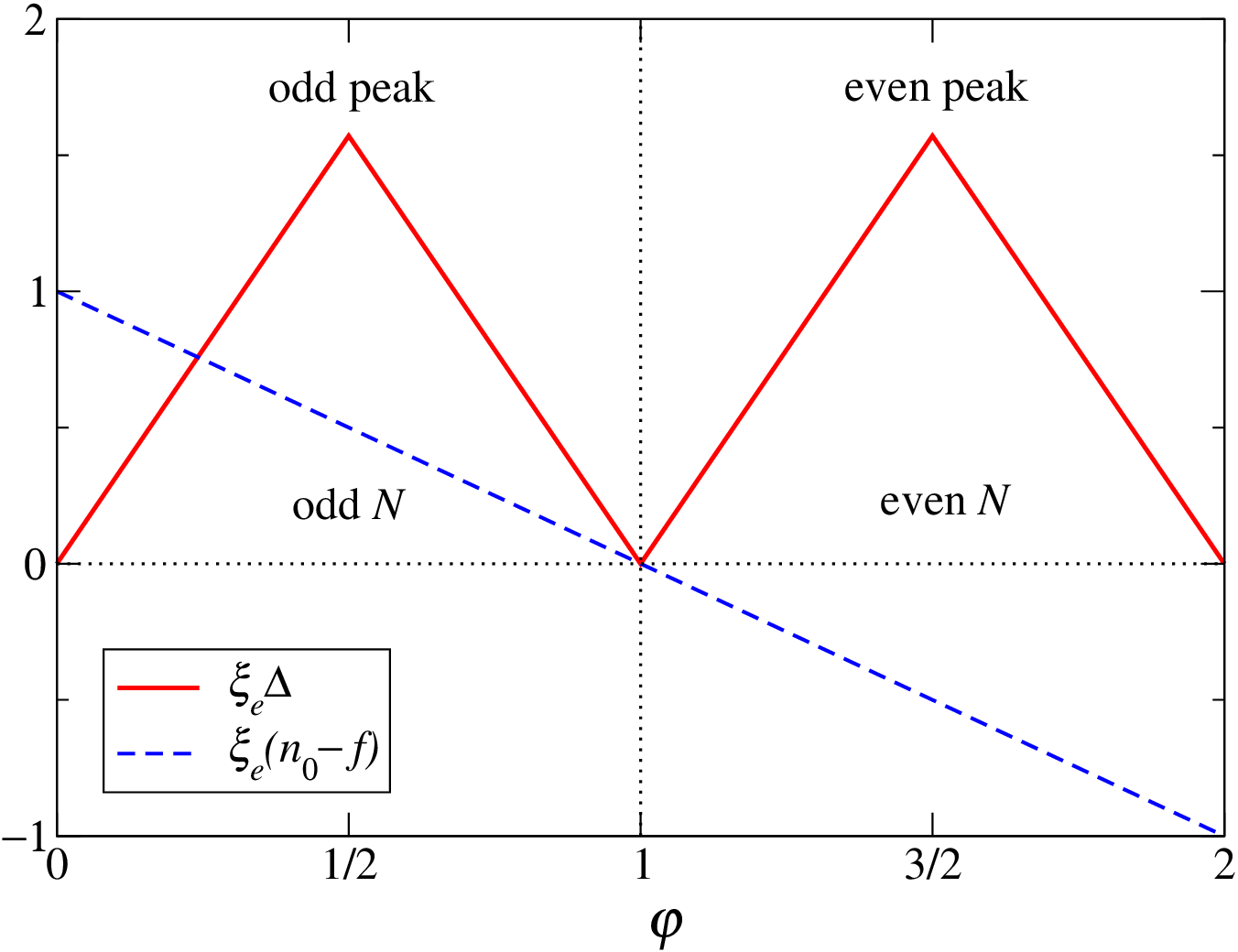}
\end{center}
\vskip-5mm
\caption{ The asymptotic behaviour of the difference between the
  lowest-energy states $\Delta$ and the particle density $\langle n_0
  \rangle$ at the center of the trap at $\mu=0$, thus $f=1/2$, and
  $T=0$. In formulae $\Delta=\pi t(\varphi)/\xi_e$ and $\langle n_0
  \rangle =(1-\varphi)/\xi_e$, where $\xi_e\sim l^{-1}$ is the
  entanglement length scale.  These results apply for any $p$,
  including $p\to\infty$.  See Refs.~\cite{CV-10-bh,CV-10-e} for
  details.  }
\label{gapdens}
\end{figure}

\begin{figure}[tbp]
\includegraphics*[scale=\graphicscale]{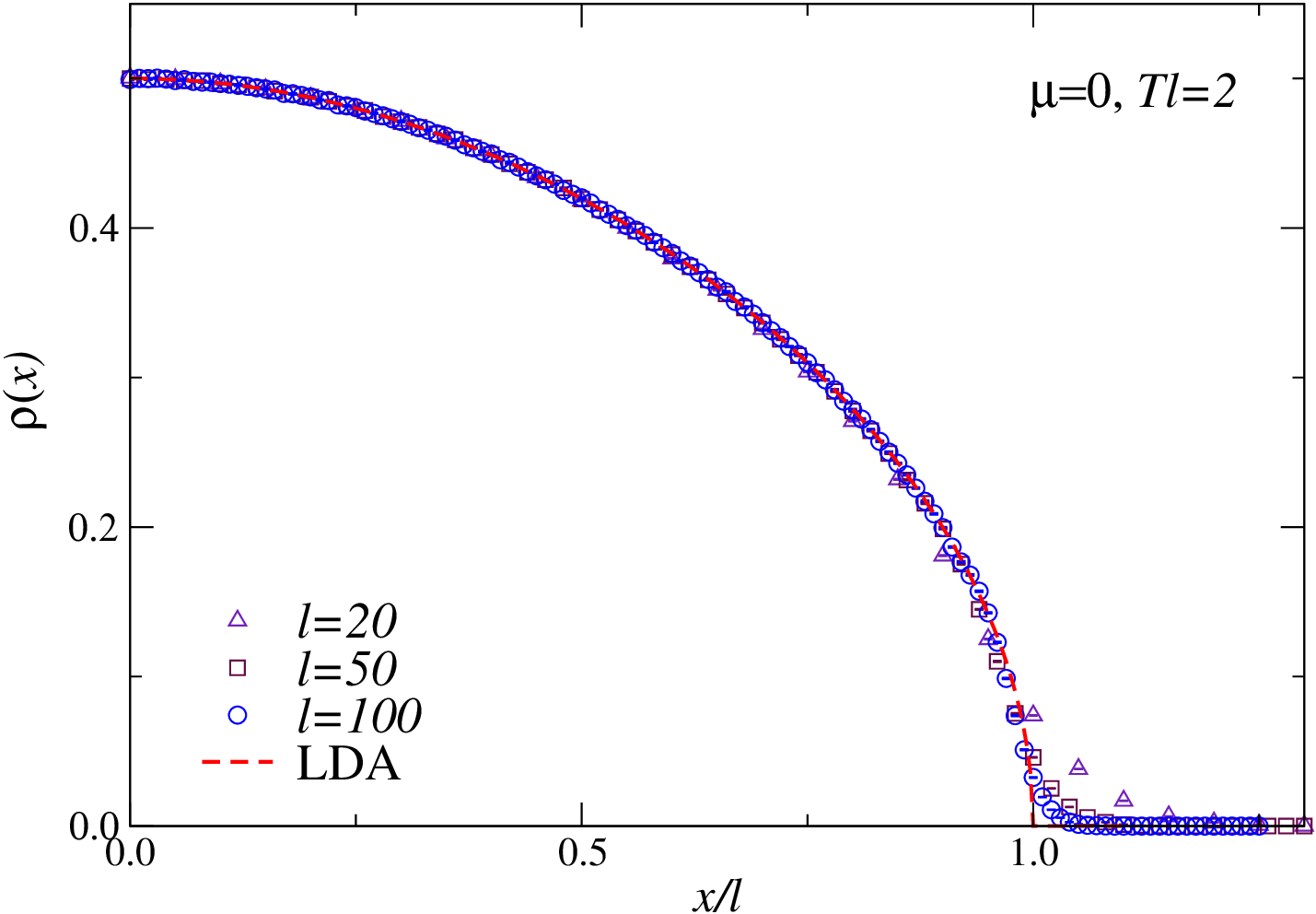}
\includegraphics*[scale=\graphicscale]{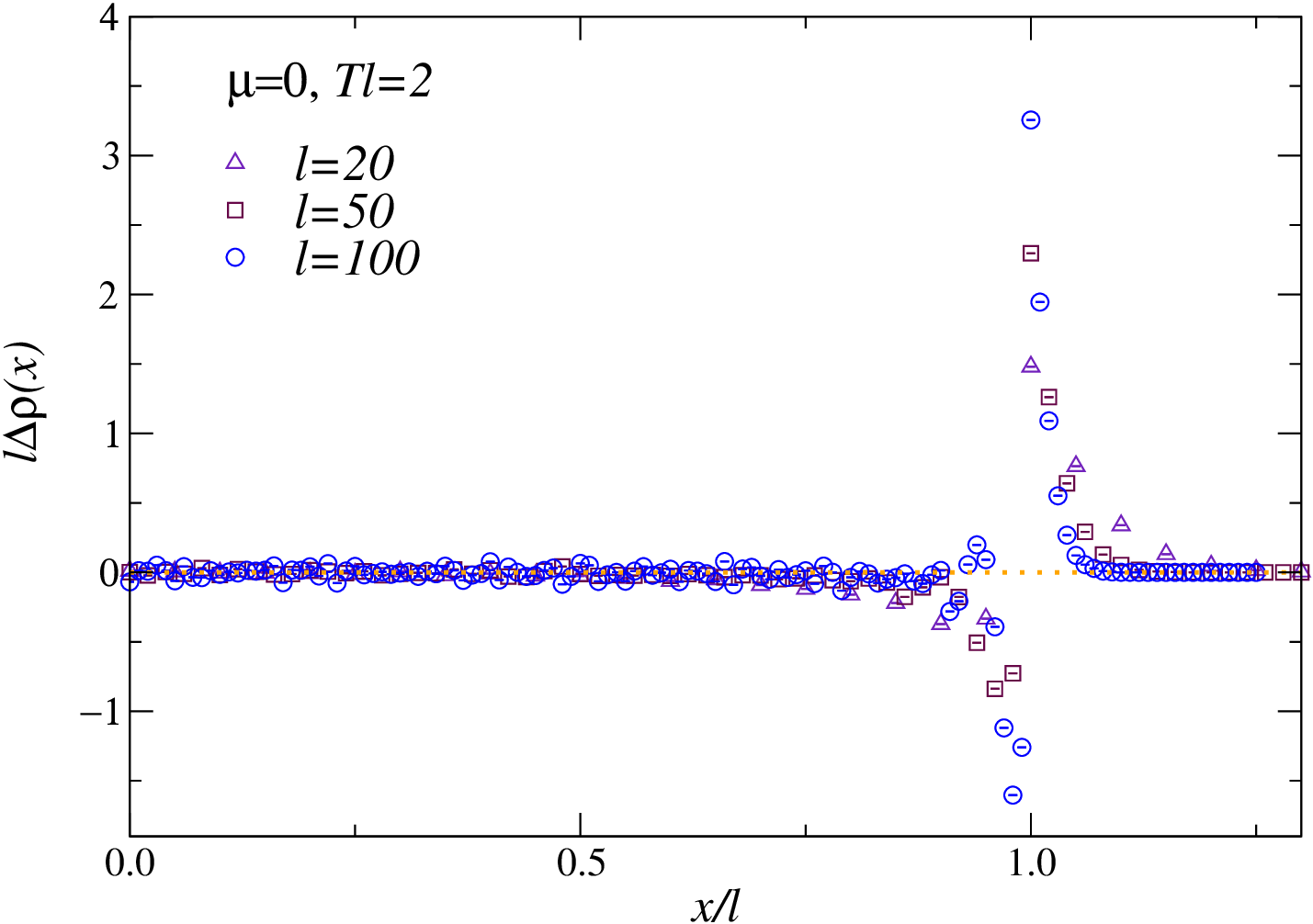}
\includegraphics*[scale=\graphicscale]{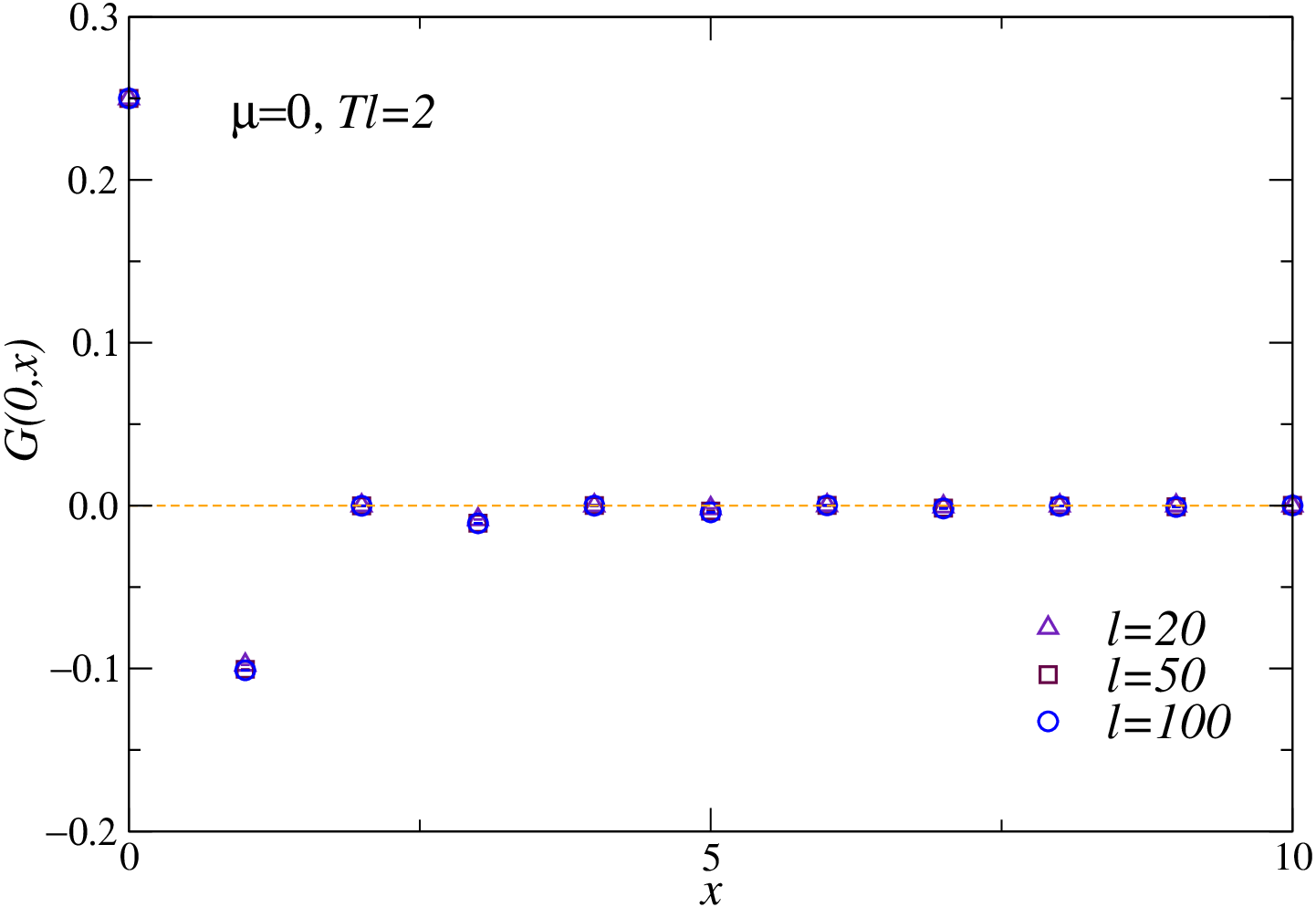}
\caption{(Color online)
We plot the particle density (top), the rescaled
subtracted particle density $l\Delta\rho$ with
$\Delta\rho\equiv\rho-\rho_{\rm lda}$ (middle),
and the particle density correlator $G(0,x)$ (bottom)
at $\mu=0$ and $T=2/l$, in the case of the harmonic potential.
}
\label{rmu0tl2}
\end{figure}

We now analyze the trap-size dependence within the gapless superfluid
phase, whose corresponding continuum theory is a conformal field
theory with $z=1$.  The trap-size dependence at zero temperature was
investigated in Refs.~\cite{CV-10-bh,CV-10-e}.  Its asymptotic
behavior turns out to be characterized by two length scales with
different power-law divergence in the large trap-size limit.  One of
them scales as $\xi\sim l$ and describes the behavior of observables
related to smooth modes, such as the half-lattice entanglement; the
other one scales as $\xi\sim l^\zeta$ with $\zeta=p/(p+1)$ and it is
found in observables involving the modes at the Fermi scale $k_F=\pi
f$, where $f$ is the filling of the homogeneous system.  Moreover, the
asymptotic power law behaviors appear modulated by periodic functions
of the trap size, which is again related to level crossings of the two
lowest states when increasing the trap size, whose interval
\begin{equation}
Q_l=l_0^{(k+1)}-l_0^{(k)} ,
\label{intql}
\end{equation}
tends to a constant: $Q_l^*=1.3110287$ for $p=2$, and $Q_l^*=1$ for
$p=\infty$. Indeed, the difference between the two lowest states
behaves asymptotically as $\Delta \sim t(\varphi)l^{-1}$, where
$\varphi$ is  defined as
\begin{equation}
\varphi = 2{l-l_0^{(2k)}\over  l_0^{(2k+2)}-l_0^{(2k)}}, 
\qquad l_0^{(2k)}\le l < l_0^{(2k+2)},
\label{varphildef}
\end{equation}
and  $t(\varphi)$ is a triangle function shown in Fig.~\ref{gapdens}
where results for $\mu=0$ are reported.

At $T=0$, the particle density and its correlation function show a
quite complicated behavior~\cite{CV-10-bh}. Indeed,
\begin{eqnarray}
&&\langle n_x\rangle \approx 
\rho_{\rm lda}(X) + l^{-1} \Re\Bigl\{h(Y,\varphi) e^{2ik_Fx} +
g(Y,\varphi)\Bigr\}, \nonumber \\
&&X = x/l, \qquad Y=x l^{-p/(p+1)}, \label{eq:rhox,sf}
\end{eqnarray}
and
\begin{eqnarray}
G_n(x,0)\approx l^{-2p/(p+1)} 
\Re\Bigl\{\tilde h(Y,\varphi) e^{2ik_Fx} + \tilde g(Y,\varphi)\Bigr\},
\label{eq:Gn,mu=0}
\end{eqnarray}
where $k_F=\pi f=\arccos\mu$, and terms suppressed by higher
powers of $l^{-1}$ are neglected.

In Fig.~\ref{rmu0tl2} we show results for the particle density and
various values of the trap size with $T=2/l$. We choose this scaling
of the temperature because the trap exponent associated with the
smooth modes is expected to be $\theta=1$~\cite{CV-10-bh}.  The
results are clearly converging toward the corresponding LDA
approximation.  Around the center of the trap, we do not observe the
modulated scaling found at $T=0$.  Only around $x\approx l$ we observe
some significant differences which decrease with increasing $l$, see
below.  Moreover, the data for the particle-density correlator
$G(0,x)$ appear to vanish after a few lattice spacings for any $l$,
without showing any particular scaling. Therefore, the leading $T=0$
scaling behavior (\ref{eq:Gn,mu=0}) associated with the exponent
$2p/(p+1)$ is also effectively averaged out by the temperature.

\section{TSS around the spatial point where $\mu_{\rm eff}(x)=1$}
\label{mueff1}

\begin{figure}[tbp]
\includegraphics*[scale=\graphicscale]{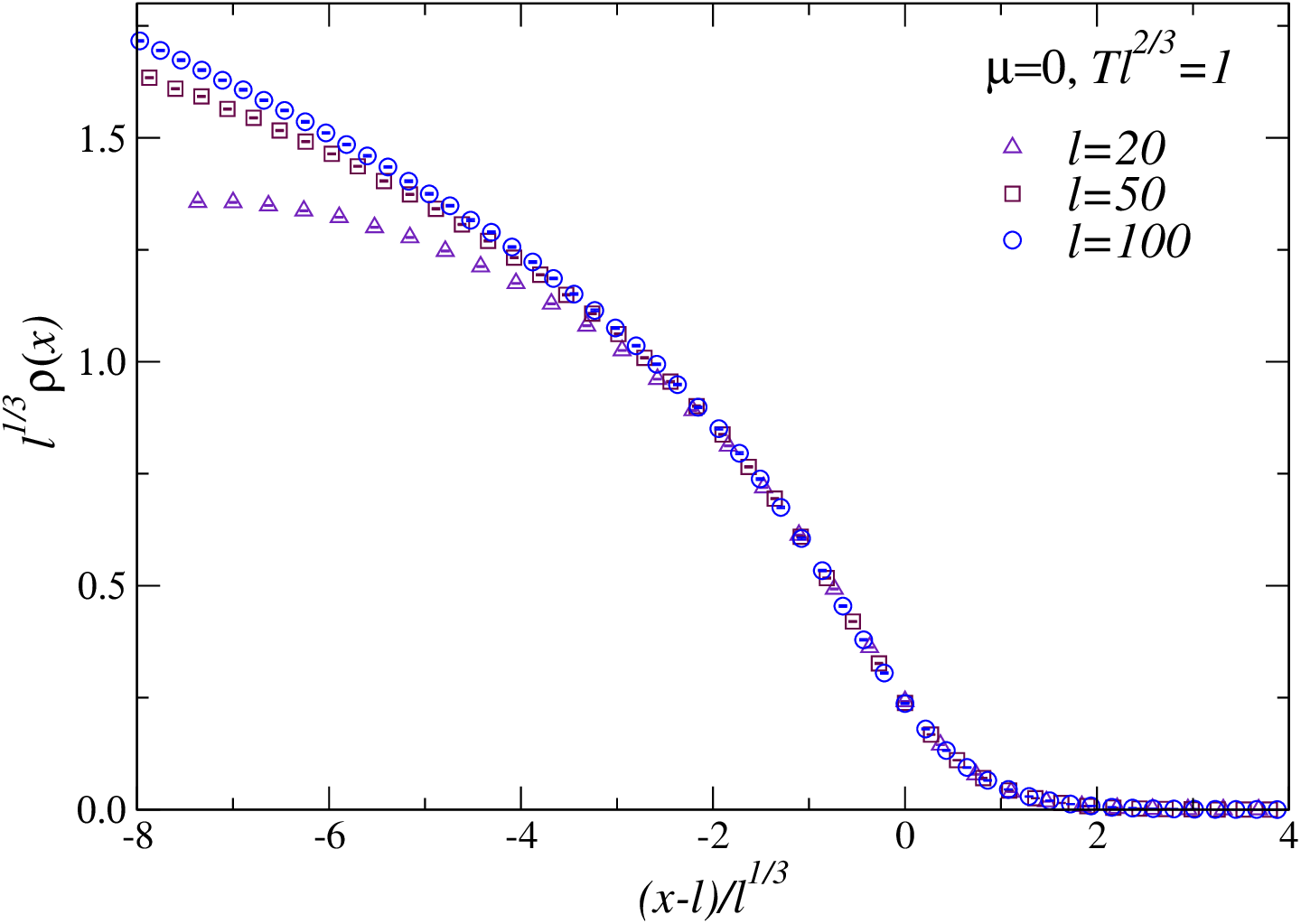}
\includegraphics*[scale=\graphicscale]{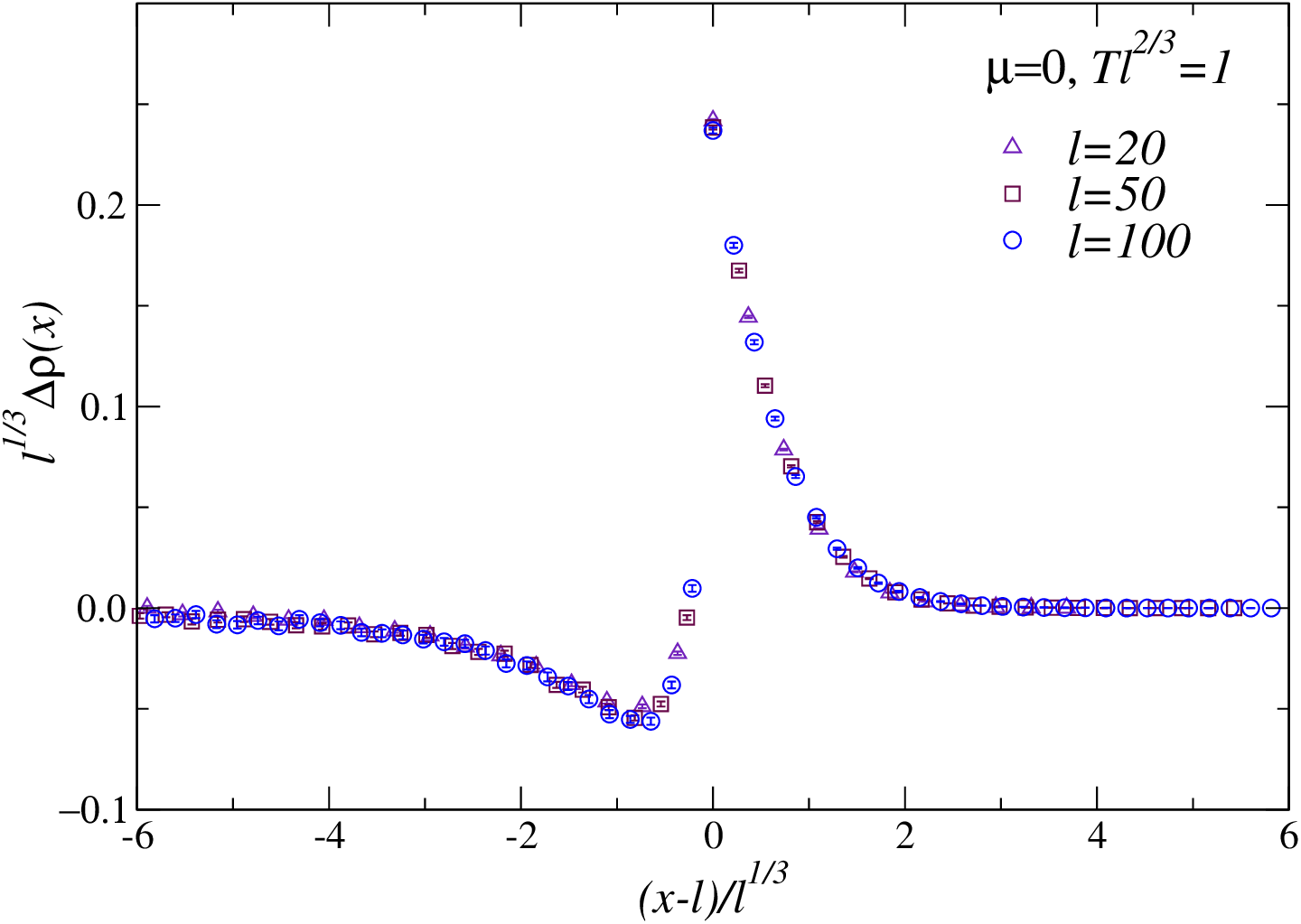}
\includegraphics*[scale=\graphicscale]{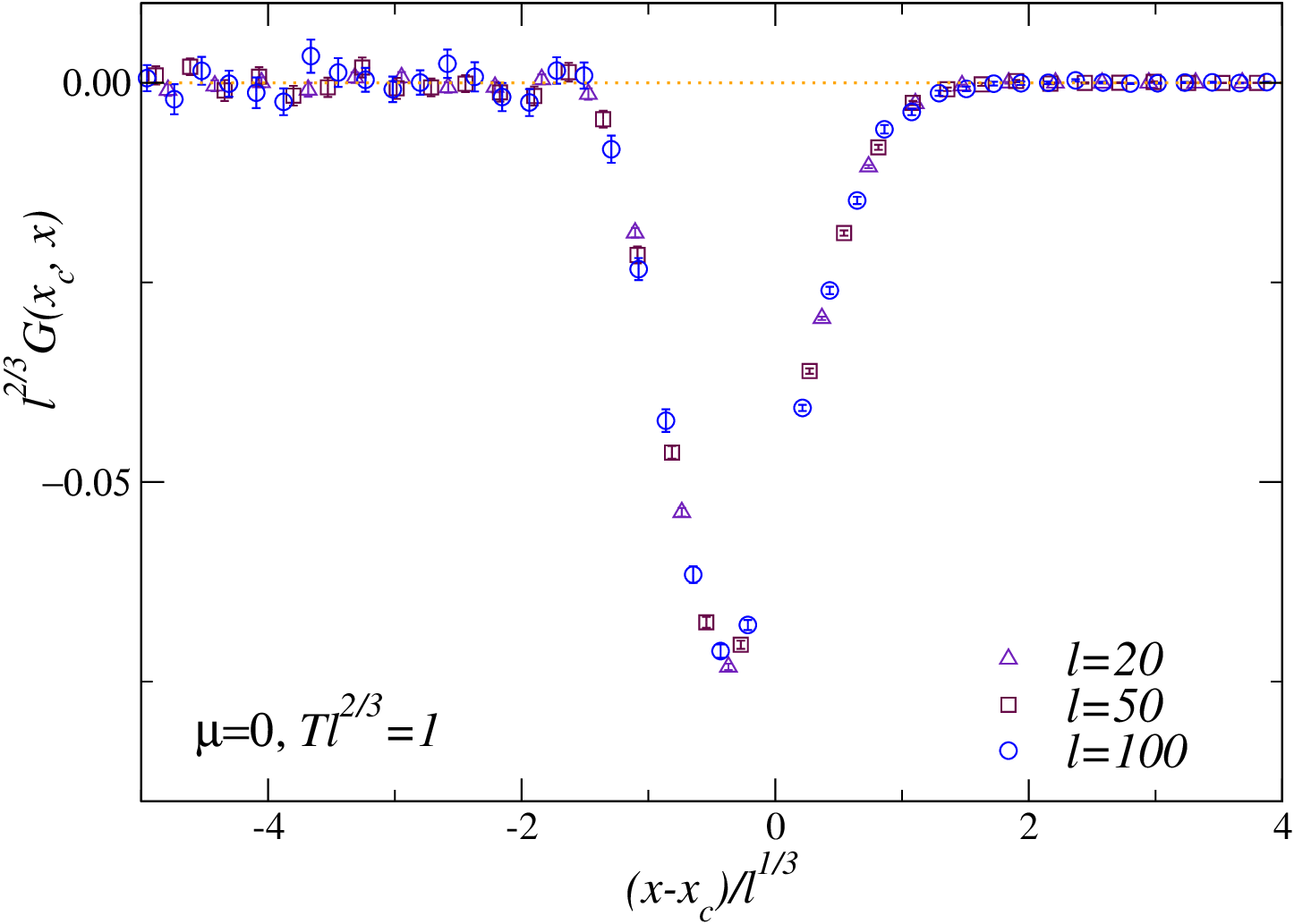}
\caption{(Color online)
Scaling of the particle density (top), the subtracted particle density
$\Delta\rho\equiv \rho-\rho_{\rm lda}$ (middle) and the particle density 
correlator (bottom) at $\mu=0$
and $Tl^{2/3}=1$ around $x=l$ where $\mu_{\rm eff}=1$.
}
\label{rmu0tl1p1xl}
\end{figure}

When $\mu<1$, the region where the particle density appears to rapidly
vanish corresponds to the region where the effective chemical
potential defined in Eq.~(\ref{mueff}) approaches the values $\mu_{\rm
eff} \approx 1$, which is the value of the chemical potential
corresponding to the transition between the superfluid phase and the
empty state, where the particle density of the ground state gets
suppressed.  We thus expect that, for generic values of $\mu$ and $p$,
the region around
\begin{equation}
x_c=l(1-\mu)^{1/p},
\label{xc}
\end{equation}
 where
$\mu_{\rm eff}(x_c)=1$, develops critical modes related to the
superfluid to vacuum transition.  The effective chemical potential can be
expanded around $x_c$ as
\begin{equation}
\mu_{\rm eff}(x) = 1 + p (1-\mu)^{(p-1)/p}{x-x_c\over l} + O[(x-x_c)^2].
\label{linV}
\end{equation}
Therefore the behavior around $x_c$ is essentially analogous to that
arising at $\mu=1$ in the presence of a linear potential $V_l\sim
r/l$.  Around $x_c$, critical modes should appear with length scale
$\xi\sim l^\sigma$, where $\sigma$ is the exponent associated with a
linear external potential, thus~\cite{CV-10-bh} $\sigma=1/3$, obtained
by replacing $p=1$ in Eq.~(\ref{thetaexp}).  We then expect that around
$x=x_c$
\begin{eqnarray}
&&\rho(x) = l^{-1/3} f_\rho[(x-x_c)/l^{1/3},Tl^{2/3}] ,\label{xc1}\\
&&\Delta\rho(x) \equiv \rho-\rho_{\rm lda}=
l^{-1/3} f_{\Delta\rho}[(x-x_c)/l^{1/3},Tl^{2/3}] ,\qquad\label{xc1b}\\
&&G(x_c,x) = l^{-2/3} f_g[(x-x_c)/l^{1/3},Tl^{2/3}] \label{xc2}.
\end{eqnarray}
In Fig.~\ref{rmu0tl1p1xl} we show  QMC results at $\mu=0$
for some values of $l$ keeping $Tl^{2/3}=1$ fixed, which
 nicely confirm the scaling behaviors (\ref{xc1}-\ref{xc2}).
  Note that analogous results are obtained for any
 $\mu<1$, around the region where $\mu_{\rm eff}(x)\approx 1$.

\section{Conclusions}
\label{conclusions}

We have investigated the interplay of temperature and trap effects in
cold particle systems at their quantum critical regime, such as cold
bosonic atoms in optical lattices at the transitions between
Mott-insulator and superfluid phases.  The theoretical framework is
provided by the 1D Bose-Hubbard model (\ref{bhm}) in the presence of
an external trapping potential, and by the TSS theory leading to the
TSS ansatz (\ref{freee}) and (\ref{gntss}).  This study is of
experimental relevance, because quasi 1D trapped particle systems have
been realized in optical lattices, see, e.g.,
Refs.~\cite{BDZ-08,PWMMFCSHB-04,KWW-05,CFFFI-09}.  Therefore,
temperature and trap effects at the Mott transitions may be
investigated by improving the accuracy of the experiments.

Our numerical study is based on QMC simulations at fixed chemical
potential and trap size.  Since the main features of the TSS at the
Mott transitions are expected to be universal with respect to the
on-site coupling $U$ (provided $U>0$), we consider the HC limit
$U\to\infty$, which is particularly convenient because it restricts
the values of the site particle number to $n_x=0,1$, and it is
expected to minimize the corrections to the universal TSS, as verified
at $T=0$~\cite{CV-10-bhn}.  We present finite-temperature results for
the particle density and the density-density correlator, at the 
quantum Mott transitions, and within the gapless superfluid phase.
Their temperature and trap-size dependences appear well described by
the simplest TSS ansatz, such as that given in Eq.~(\ref{gntss}) with
$\theta=p/(p+2)$. In particular, the finite-temperature data do not
show evidence of the periodic modulations of the asymptotic behaviors
which are observed at $T=0$~\cite{CV-10-bh}. They are effectively
averaged out by a nonzero temperature, leaving only a plain power-law
TSS behavior.

\acknowledgements The QMC simulations were performed at the INFN Pisa
GRID DATA center, using also the cluster CSN4.

\appendix

\section{The QMC simulations}
\label{qmcsim}

Our numerical study of the BH model (\ref{bhm}) is based on QMC
simulations using the stochastic series expansion
method~\cite{Sandvik-99} with the directed loop
algorithm~\cite{SS-02}.  We compute the particle density $\langle n_x
\rangle$ and its connected correlation function $\langle n_x n_y
\rangle_c$.

The QMC simulations are performed at fixed temperature $T$, chemical
potential $\mu$, trap size $l$, and lattice size $L$ with open
boundary conditions.  In order to avoid finite-size effects, we choose
$L$ sufficiently large to effectively obtain $L\to\infty$ data within
the statistical errors.  

QMC simulations require also to fix some algorithm parameters, such as
the order $N_{\rm tr}$ of the truncation of the Taylor expansion of
the partition function, and the number of operator loops $N_{\rm{loops}}$.
We set $N_{\rm tr}$ in the standard way as $N_{\rm tr}= 1.5 \, M_{\rm
max} \, N_{\rm{bonds}}/T$ where $M_{\rm max}$ is the highest matrix
element of the single bond Hamiltonians and $N_{\rm{bonds}}$ is the
number of interacting site pairs. At the end of each simulation we
analyze the fluctuation in the order of the series expansion, and
verify whether the averaged order is safely less than $N_{\rm tr}$,
and its deviations have the expected behavior proportional to the
square root of the mean value.  We define a MC step as a single sweep
of identity and diagonal operators in the expansion of the partition
function followed by a number $N_{\rm{loops}}$ of directed operator
loop updates.  The number $N_{\rm{loops}}$ is fixed by requiring that
the number of visited vertices is globally of the order of $N_{\rm
tr}$ in a MC step.  

Typical statistics of our QMC simulations are a few million MC steps
for each set of parameters.  We have checked that, in all QMC
simulations presented here, the autocorrelation times of the
observables that we compute are smaller than $10^2$ in units of MC
steps.  Effective $L\to\infty$ data within the resulting statistical
errors are obtained by taking $L/l\approx 4$ for the simulations at
$\mu=1$ and $\mu=0$, and $L/l\approx 9$ for $\mu=-1$.

Let us now discuss the operator-loop equations. Since the Hamiltonian
of the HC BH model can be written as a sum of bond terms, and since
there is a one-to-one correspondence between the site occupation
number and the states of the XXZ model~\cite{xxmodel}, the non-zero
matrix elements are the same as those reported in Ref.~\cite{SS-02}
(see its Sec. 2). Of course, their values are different, in particular
the diagonal matrix elements are site dependent, due to the existence
of the trap which gives rise to a space-dependent effective chemical
potential $\mu_{\rm eff}(x)$.  Using considerations based on the
symmetry and conservation laws, we can divide the transition
probabilities involved in the operator-loop update in eight
independent groups for each site, as in Ref.~\cite{SS-02} (see its
Sec. 3), with three possible bounces (probabilities to exit the actual
visited vertex of the operator loop from the same entrance leg) for
each group.  The corresponding system of equations are solved by
minimizing the bounces.  As a result, we allow one bounce within each
group in the QMC simulations at $\mu =1$ and $\mu =0$, and two bounces
at $\mu =-1$.


\begin{thebibliography}{99}

\bibitem{BDZ-08}
I. Bloch, J. Dalibard, and W. Zwerger,
Rev.\ Mod.\ Phys.\ 80, 885 (2008).

\bibitem{GBMHS-02}
M. Greiner,
I. Bloch, M.O. Mandell, T. H\"ansch, and T. Esslinger,
Nature 415, 39 (2002).

\bibitem{SMSKE-04}
T. St\"oferle, H. Moritz, C. Schori, M. K\"ohl, and T. Esslinger,
Phys.\ Rev.\ Lett.\ 92, 130403 (2004).

\bibitem{PWMMFCSHB-04} B. Paredes, A. Widera, V. Murg, O. Mandel, S.
  F\"olling, I, Cirac, G. Shlyapnikov, R.W. H\"ansch, and I. Bloch, Nature 429,
  277 (2004).

\bibitem{KWW-05}
T. Kinoshita, T. Wenger, and D.S. Weiss, Science 305, 1125 (2004);
Phys.\ Rev.\ Lett.\  95, 190406 (2005).

\bibitem{HSBBD-06}
Z. Hadzibabic,
P. Kr\"uger, M. Cheneau, B. Battelier, and J. Dalibard,
Nature 441, 1118 (2006).

\bibitem{FWMGB-06}
S. F\"olling, A. Widera, T. M\"uller, F. Gerbier, and I. Bloch,
Phys.\ Rev.\ Lett.\ 97, 060403 (2006).

\bibitem{SPP-07}
I.B. Spielman, W.D. Phillips, and J.V. Porto,
Phys.\ Rev.\ Lett.\ 98, 080404 (2007);
Phys.\ Rev.\ Lett.\ 100, 120402 (2008).

\bibitem{CFFFI-09} D. Cl\'ement, N. Fabbri, L. Fallani, C. Fort, and
  M. Inguscio, 
  Phys.\ Rev.\ Lett.\ 102, 155301 (2009).

\bibitem{JBCGZ-98} D. Jaksch, C. Bruder, J.I. Cirac, C.W. Gardiner,
  and P. Zoller, 
  Phys.\ Rev.\ Lett.\ 81, 3108 (1998).

\bibitem{FWGF-89}
M.P.A. Fisher, P.B. Weichmann, G. Grinstein, and D.S. Fisher,
Phys.\ Rev.\ B 40, 546 (1989).

\bibitem{footnote-2d}
In two dimensions logarithmic corrections may arise
due to the fact that $d=2$ is the upper critical space dimension
of the corresponding continuum theory.


\bibitem{BRSRMDT-02}
G.G. Batrouni, V. Rousseau, R.T. Scalettar, M. Rigol, A. Muramatsu,
P.J.H. Denteneer, and M. Troyer,
Phys.\ Rev.\ Lett.\ 89, 117203 (2002).

\bibitem{KPS-02}
V.A. Kashurnikov, N.V. Prokofev, and B.V. Svistunov, 
Phys.\ Rev.\ A 66, 031601 (2002).

\bibitem{KSDZ-04}
C. Kollath, U. Schollw\"ock, J. von Deft, W. Zwerger,
Phys.\ Rev.\ A 69, 031601 (2004).

\bibitem{PRHD-04}
L. Pollet, S. Rombouts, K. Heyde, and J. Dukelsky,
Phys.\ Rev.\ A 69, 043601 (2004). 

\bibitem{WATB-04}
S. Wessel, F. Alet, M. Troyer, and G.G. Batrouni,
Phys.\ Rev.\ A 70, 053615 (2004).

\bibitem{RM-04}
M. Rigol and A. Muramatsu, Phys.\ Rev.\ A 70, 031603 (2004);
Phys.\ Rev.\ A 72, 013604 (2005).

\bibitem{DLVW-05}
B. DeMarco, C. Lannert, S. Vishveshwara, and T.-C. Wei,
Phys.\ Rev.\ A 71, 063601 (2005).

\bibitem{GKTWB-06}
O. Gygi, H.G. Katzgraber, M. Troyer, S. Wessel,  and G.G. Batrouni,
Phys.\ Rev.\ A 73, 063606 (2006).

\bibitem{ULR-06}
L. Urba, E. Lundh, and A. Rosengren, J. Phys.\ B 39, 5187 (2006).

\bibitem{RBRS-09}
M. Rigol, G.G. Batrouni, V.G. Rousseau and R.T. Scalettar,
Phys.\ Rev.\ A 79, 053605 (2009).

\bibitem{CV-10} M. Campostrini and E. Vicari, 
Phys.\ Rev.\ A 81,  023606 (2010).

\bibitem{CV-09} M. Campostrini and E. Vicari, Phys.\ Rev.\ Lett.\ 102, 240601
  (2009).

\bibitem{CV-10-bh} M. Campostrini and E. Vicari, Phys.\ Rev.\ A 81,
  063614 (2010).

\bibitem{QSS-10} S.L.A.~de~Queiroz, R.R.~dos~Santos, and R.B.~Stinchcombe,
Phys. Rev. E 81, 051122 (2010).

\bibitem{Sandvik-99}
A.W. Sandvik, Phys. Rev.  B 59, R14157 (1999).

\bibitem{SS-02}
O.F. Syljuasen and A.W. Sandvik,
Phys. Rev. E 66,  046701 (2002).

\bibitem{xxmodel} In the HC limit the 1D BH model can be mapped into
the XX chain model with a space-dependent transverse external field,
$H_{\rm XX} = - J \sum_i \left( S^x_i S^x_{i+1} + S^y_i S^y_{i+1}
\right)+ \sum_i [\mu+V(x_i)] S^z_i$, where $S^a_i=\sigma^a_i/2$ and
$\sigma^a$ are the Pauli matrices, which are related to the boson
operators $b_i$ by $\sigma^x_i = b_i^\dagger + b_i$, $\sigma^y_i =
i(b_i^\dagger - b_i)$, $\sigma^z_i = 1-2b_i^\dagger b_i$.  
Then, by a Jordan-Wigner transformation, one
can further map it into a model of spinless fermions, see, e.g.,
Ref.~\cite{Sachdev-book}.

\bibitem{CV-10-bhn} M. Campostrini and E. Vicari, Phys.\ Rev.\ A 82,
  063636 (2010).

\bibitem{Sachdev-book}
S. Sachdev, {\em Quantum Phase Transitions}
(Cambridge Univ.\ Press, 1999).



\bibitem{CV-10-e} M. Campostrini and E. Vicari, 
J. Stat. Mech.: Theory Exp.  (2010) P08020.




\end{thebibliography}
\end{document}